\documentclass[nofootinbib]{revtex4-2}
\usepackage{bbold}
\usepackage{siunitx}
\usepackage{gensymb}
\usepackage{mathtools}
\usepackage[colorlinks=true,linkcolor=black]{hyperref}
\usepackage[left=3cm, right=3cm]{geometry}

\newcommand{\beq}{\begin{equation}}
\newcommand{\eeq}{\end{equation}}

\newcommand{\irm}{{\rm i}}

\newcommand\AddLabel[1]{%
  \refstepcounter{equation}
  (\theequation)
  \label{#1}
}

\begin{document}
\title{Opportunities and limits of lunar gravitational-wave detection}

\author{Andrea Cozzumbo} 

\author{Benedetta Mestichelli}
\author{Marco Mirabile}
\author{Lavinia Paiella}
\author{Jacopo Tissino}
\author{Jan Harms}
\email[]{jan.harms@gssi.it}

\affiliation{Gran Sasso Science Institute (GSSI), I-67100 L'Aquila, Italy\\
INFN, Laboratori Nazionali del Gran Sasso, I-67100 Assergi, Italy}






\begin{abstract}
A new era of lunar exploration has begun with participation of all major space agencies. This activity brings opportunities for revolutionary science experiments and observatories on the Moon. The idea of a lunar gravitational-wave detector was already proposed during the Apollo program. The key characteristic of the Moon is that it is seismically extremely quiet. It was also pointed out that the permanently shadowed regions at the lunar poles provide ideal conditions for gravitational-wave detection. In recent years, three different detector concepts were proposed with varying levels of technological complexity and science potential. In this paper, we confront the three concepts in terms of their observational capabilities based on a first more detailed modeling of instrumental noise. We identify important technological challenges and potential show-stoppers.
\end{abstract}

\maketitle

\section{Introduction}
Lunar gravitational-wave (GW) detection was originally proposed by Joseph Weber who was the lead scientist behind all the early efforts to build gravitational-wave detectors. His Lunar Surface Gravimeter was deployed on the Moon by the crew of the Apollo 17 mission in December 1972 \cite{GiEA1977}. The main science goal was to observe surface vibrations of the Moon caused by passing GWs. The instrument did not perform as expected, and today we also know that the sensitivity would not have been enough to detect GW signals. Nonetheless, the idea was valid since the Moon was already known at that time to be seismically extremely quiet \cite{NaEA1981}. Models of the lunar seismic background indicate that vibration amplitudes in the frequency band suitable for GW detection might be several orders of magnitude weaker than on Earth \cite{LoEA2009}. Furthermore, the permanently shadowed regions (PSRs) at the lunar poles provide thermally stable conditions, and they might very well be the coldest regions in our solar system \cite{WiEA2019}.

Following opportunities to submit mission ideas to NASA and ESA calls, three new concepts were proposed in 2020: the Lunar GW Antenna (LGWA) \cite{HaEA2021a}, the Lunar Seismic and Gravitational Antenna (LSGA) \cite{KaEA2020}, and the Gravitational-wave Lunar Observatory for Cosmology (GLOC) \cite{JaLo2021} soon followed by a study of a concept similar to GLOC called LION \cite{ASEA2021}. Another concept similar to LGWA was recently proposed using a different sensor technology \cite{LiEA2023}. The LGWA and LSGA concepts exploit the response of the Moon to GWs, while GLOC and LION work like the Virgo/LIGO detectors on Earth \cite{AcEA2015,LSC2015}. In the following, we will refer to the GLOC/LION type concepts as long-baseline interferometers with suspended test masses (LBI-SUS), and to LSGA type concepts as long-baseline interferometers with ground optics (LBI-GND). 

A recent study found that sensitivity limits coming from the seismic background create a separation of the observation bands of the three concepts \cite{Har2022a}. The order of concepts from low to high observation frequencies is LBI-GND, LGWA, and LBI-SUS. As we will show in this paper, none of the concepts can ever reach good sensitivity below 1\,mHz, which must be left to space-based detectors like LISA \cite{ASEA2017}, and there is no strong motivation to observe GWs above a few Hertz with lunar detectors since this is where terrestrial detectors can already achieve an excellent sensitivity \cite{ET2020}. Inside this designated band of lunar GW detection from 1\,mHz to a few Hertz lies the decihertz band, which plays a strategic role for GW science. A decihertz detector would be the \emph{missing link} between the observation bands of LISA and terrestrial detectors. The decihertz band offers immense opportunities for breakthrough science \cite{CuHo2009,MSV2018,JSC2019,SeEA2020,HaEA2021a,JaLo2021}. If the sensitivity targets of proposed space-based decihertz detectors like the Big Bang Observer \cite{Phi2003} and DECIGO \cite{KaEA2021} can be reached, they would be able to observe primordial GWs even when assuming conservative slow-roll inflationary models \cite{CuHa2006,HaEA2008}. 

In this paper, we compare sensitivity models and observational capabilities of the three lunar GW detector concepts. So far, only the LGWA concept was presented with a detailed noise budget \cite{vHEA2023}. We present a first noise model for an LBI-GND concept and a more detailed analysis of instrumental noise of a LBI-SUS concept. In contrast to the sensitivity targets proposed for GLOC and LION, we believe that such a detector needs to be more ambitious and cover the entire decihertz band to justify the immense cost and effort to build such a detector on the Moon. Proposed next-generation terrestrial detectors like Einstein Telescope \cite{ET2020} or Cosmic Explorer \cite{EvEA2021} will already reach down to a few Hertz with GW observations. In section \ref{sec:acc}, we briefly review the LGWA concept and noise model. A more detailed analysis of instrument-noise models is provided in section \ref{sec:noise} for long-baseline laser-interferometer concepts. We use these models to calculate noise budgets for LBI-GND, LBI-SUS concepts, which we present in section \ref{sec:ifo}. We then take the sensitivity models and compare observational capabilities of all three concepts in section \ref{sec:capa}.

\section{Inertial acceleration measurement}
\label{sec:acc}
The LGWA was proposed as an array of 4 seismometers deployed in a PSR. Each seismometer reads out the lunar surface displacement produced by GWs with respect to an inertial reference mass. The array helps to distinguish between the seismic background and GW signals. The array diameter depends on the correlation length of the seismic field around 0.3\,Hz, where LGWA is designed to have its peak sensitivity. Correlations between sensors must be high enough to coherently analyze the seismic background, but the sensors cannot be too close either since differential effects between sensors strongly boost the efficiency of background reduction methods. It is expected that the optimal diameter is about a kilometer, which is quite small and due to the fact that the lunar regolith and the underlying geology do not have the same solid structure of the typical terrestrial geology. The requirement of 4 seismic stations comes from the idea that three stations are needed to uniquely identify propagation directions along the surface, and to use this information to predict (and then subtract) the seismic disturbance at a fourth station. Deployment inside a PSR is crucial to profit from low temperatures and better thermal stability. Temperature changes were responsible for excess noise especially in horizontal seismic channels of the Apollo seismometers \cite{DuSu1974}.

Each LGWA station will have two horizontal seismic channels for vibration measurements between 1\,mHz and 10\,Hz. This means that LGWA has a total of 8 GW channels. The baseline design of the LGWA payload consists of a compact niobium Watt's linkage with a resonance frequency of 0.25\,Hz and an inertial reference mass of 10\,kg \cite{vHEA2023}. A Watt's linkage is a combination of a pendulum and an inverted pendulum allowing for a very low resonance frequency in a compact device \cite{vHEA2023,BERTOLINI2006616}; the specific design, which permits the fabrication of a (quasi) monolithic mechanical structure \cite{BeEA2006,Ferreira_2021}, is adapted from an earlier prototype. The main sensitivity limitation below 0.5\,Hz comes from suspension thermal noise connected to structural damping of the niobium mechanical structure. While niobium quality factors of $5\cdot10^7$ were observed in the past at low temperatures \cite{BiEA2002}, a conservative estimate of $Q=10^4$ is chosen for the LGWA baseline design since it is expected that the electro-discharge machining especially of the sub-mm thick flexures of the Watt's linkage might degrade its mechanical quality. Displacements relative to the inertial reference mass will be read out with a compact laser interferometer. The readout noise will limit the sensitivity of the sensor above 0.5\,Hz. The mechanical structure will be cooled down from <50\,K ambient temperatures of the PSR to 4\,K with a sorption cooler, which permits the use of superconducting coil actuators.

\begin{figure}
\centering{\includegraphics[width=0.8\textwidth]{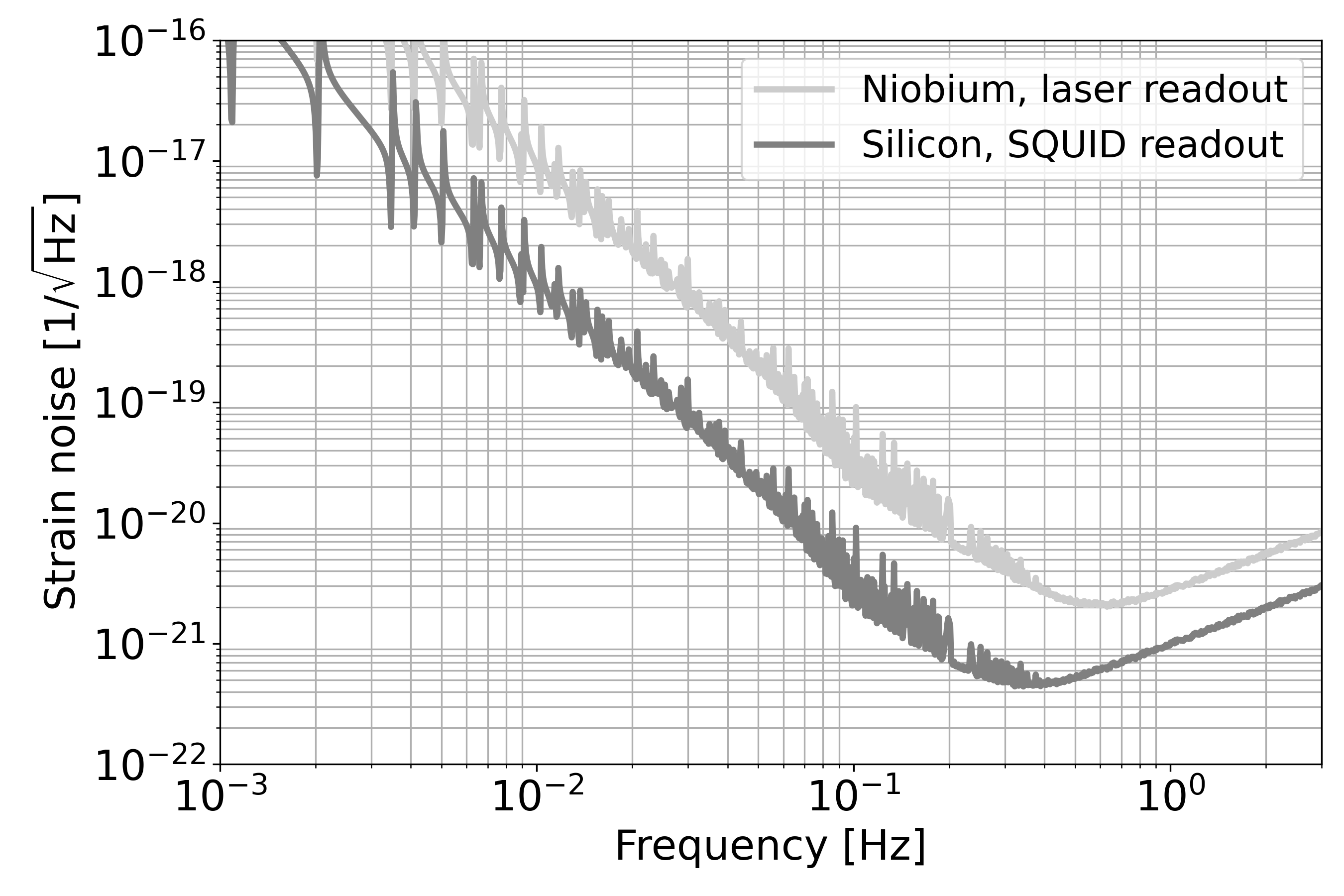}}
\caption{Sensitivity targets for a single LGWA seismic channel (the full detector has 8 seismic channels). The sensitivity below 0.3\,Hz is limited by suspension thermal noise, where the niobium model assumes a quality factor $Q=10^4$ and the silicon model $Q=10^6$. The sensitivity above 0.3\,Hz is limited by readout noise assuming that the seismic background can be reduced sufficiently. A detailed noise budget was presented in \cite{vHEA2023}.}
\label{fig:lgwasens}
\end{figure}
An alternative sensor concept is under investigation, where the niobium material is substituted by silicon, and the laser-interferometric readout by a magnetic readout using superconducting coils and superconducting quantum-interference devices (SQUID) as amplifiers \cite{vHEA2023}. This concept would lead to sensitivity improvements over the entire observation band since it reduces suspension thermal noise (assuming a quality factor of $Q=10^6$ for silicon) and readout noise (sub-femtometer) with respect to the baseline design. It would also reduce power consumption of the payload, which might be a strong asset at a deployment location where solar panels cannot easily be used to deliver energy. It should be noted though that even with niobium and laser-interferometric readout, it might be possible to reach this performance \cite{BiEA2002,EcGe2022}. The two sensitivity models are shown in figure \ref{fig:lgwasens}.

The lunar GW response model is another important ingredient for the evaluation of the detector sensitivity curve. The response below the decihertz band is best modeled by normal-mode simulations \cite{Ben1983,CoHa2014b, HaEA2021a}. Above 0.1\,Hz, topography and regional geology are expected to play an important role, because seismic waves above 0.1\,Hz are short enough to show significant interaction with these heterogeneous structures. More detailed simulations of such a system with appropriate tools are under preparation. For the time being, we match the decihertz response to known features. For example, starting from the simplified Dyson response, which states that the effective detector baseline of a homogeneous ground to GWs is the shear-wave length divided by $\pi$, we include a correction from resonant amplification of the ground response. The quality factor of ground material even at relatively shallow depth is in the thousands \cite{GaEA2019}. Generally, the rock quality is only an indication of the achievable amplification factor, and the actual energy loss can be due to coupling into deep parts of the lunar interior and scattering. However, measurements of, e.g., decay times of moonquake coda (the diffusive waveform at the end of a moonquake) are consistent with quality factors in the thousands, which indicates an actual storage capability of seismic energy \cite{LaEA1970}. This would also mean an enhanced GW response. However, due to the strong scattering of the waves, at least when propagating over larger distances, it is unclear how much the amplification can be exploited to improve sensitivity to GW signals. As a conservative estimate, we assume amplification factors of a few 100 in the decihertz band. The LGWA sensitivity estimate could easily be off in either direction by an order of magnitude.

\section{Instrument noise of long-baseline laser interferometers}
\label{sec:noise}

\subsection{Quantum noise}
For the modeling of quantum noise, we will follow the formalism initially developed for terrestrial GW detectors \cite{BuCh2001,BHS2018}. Important here is to distinguish between interferometers with suspended test masses and interferometers where the test masses are mounted to the ground. In the former case, there is additional noise at low frequencies caused by the quantum radiation-pressure fluctuations of the laser beam. The other important distinction is between a laser interferometer that has arm cavities and one that does not. Arm cavities are optical resonators formed by a pair of mirrors, which resonantly amplify the light power and signal response. A detailed analysis whether arm cavities are required are beyond the scope of this paper. We choose a system without arm cavities for LBI-GND, and a system with arm cavities for LBI-SUS. Arm cavities add complexity to the system, but they give an important handle to control and reduce noise couplings from auxiliary degrees of freedom, which might prove crucial for the ambitious LBI-SUS concept.

The quantum noise of an LBI-GND detector is given by
\beq
h_{\rm QN}^{\rm gnd}(f) = \sqrt{\frac{\hbar\omega_0}{P}}\frac{2c}{\omega_0L},
\label{eq:sensgnd}
\eeq
where $L$ is the length of the interferometer arm, $\omega_0$ is the angular frequency of the laser, $c$ is the speed of light, $P$ is the power of the laser beam in each arm, and $\hbar$ is Planck's constant. There is no significant radiation-pressure noise in this case since the optics are mounted to ground. If instead we consider arm cavities (optical resonators) and suspended test masses, then a set of transfer functions need to be considered, which describe how the signal and noise fields propagate inside the detector:\\[0.5cm]
\begin{tabular}{lp{6.5cm}r}
$\displaystyle \mathcal R(f) = \exp(4\pi\irm Lf/c)
\begin{pmatrix}
1 & 0 \\
-\mathcal K(f) & 1
\end{pmatrix}$ & Round-trip propagation of a field inside an optical resonator of a detector arm (in quadrature formalism \cite{BHS2018}) & \\
 & & \\
$\displaystyle \mathcal C(f) = (\mathcal R(f)-\rho\mathbb 1)\cdot(\mathbb 1-\rho\mathcal R(f))^{-1}$ & Reflection of optical field from optical resonator & \\
 & & \\
$\displaystyle \mathcal T(f) = \tau(\mathbb 1-\rho\mathcal R(f))^{-1}$ & Transmission of field into or out of the optical resonator & \\
 & & \AddLabel{eq:quantum} \\
$\displaystyle S(f) = \vec\eta^\top\cdot\Re(\mathcal C(f)\cdot\mathcal C(f)^\dagger)\cdot\vec\eta$ & Power-spectral density of quantum noise at the detector output; physical units are accounted for in equation (\ref{eq:sensgnd}) &\\
 & & \\
$\displaystyle r(f) = \vec\eta^\top\cdot\mathcal T(f)\cdot 
\begin{pmatrix}
0 \\
1
\end{pmatrix}$ & Signal field at the detector output (in quadrature formalism \cite{BHS2018}); physical units are accounted for in equation (\ref{eq:sensgnd})  & \\
 & & \\
$\displaystyle h_{\rm QN}^{\rm sus}(f) = h_{\rm QN}^{\rm gnd}(f)\frac{\sqrt{S(f)}}{r(f)}$ & Sensitivity limitation caused by quantum noise in units of GW strain &
\end{tabular}\\[0.5cm]
Here, $\mathcal K(f) = 8\omega_0P/(mc^2(2\pi f)^2)$, $m$ is the suspended test mass, and $\vec \eta$ is a unit vector that specifies the linear combination of amplitude and phase quadrature observed by the homodyne detector \cite{BHS2018}. Here we choose $\vec \eta=(0,1)$, which represents a measurement of the phase quadrature. The two arm cavities are assumed to be loss free except for the transmissivity $\tau=\sqrt{1-\rho^2}$ of their input mirrors. The unit vector $(0,1)$ is chosen to minimize quantum noise at most frequencies, but it can in principle have other directions controlled by the homodyne detection system.

\subsection{Model of the ambient seismic field}
In order to model seismic noise in LBI-GND and LBI-SUS detectors, one first needs a model of ground vibrations. In general, not only the seismic displacement spectrum matters, but also ground tilt and spatial correlations. For the displacement spectra, we will use the study \cite{LoEA2009} by Lognonné et al, where time series of ground vibrations caused by meteoroid impacts were simulated. The simulations predicted a root-mean-square (rms) below 75\,fm 50\% of the time and below 0.5\,pm 90\% of the time. We would like to convert the rms into a power spectral density (PSD), which requires an assumption of the frequency dependence of the spectrum. We do not know at present what the background spectrum looks like. Models of meteoroid impacts indicate that it might not be approximated well by a simple power law \cite{DaEA2018}. Nevertheless, in this study, we adopt a simple model where the seismic displacement spectrum is proportional to $1/f^{1/2}$. Considering that the rms in Lognonné et al was evaluated down to 17\,mHz, the noise spectrum consistent with an rms of 75\,fm is
\beq
\xi(f)=\sqrt{\frac{1\,{\rm Hz}}{f}}10^{-14}\,{\rm m/Hz^{1/2}}.
\label{eq:seisspectrum}
\eeq
It should be noted that this model predicts a background noise that is higher than the fm/Hz$^{1/2}$ background on low-order normal-mode resonances estimated by Harms et al \cite{HaEA2021a}. It is likely that an extrapolation of the PSD derived from simulations in the decihertz band towards lower frequencies is inaccurate. The main background in the decihertz band is produced by the impact of many very small meteoroids and these small impacts can hardly excite the global quadrupole modes at mHz frequencies. Additional numerical simulations are required, and upcoming lunar seismic experiments like the Farside Seismic Suite \cite{PaEA2022} will shed new light on the distribution and properties of seismic events.

\begin{figure}
\centering{
\includegraphics[width=0.48\textwidth]{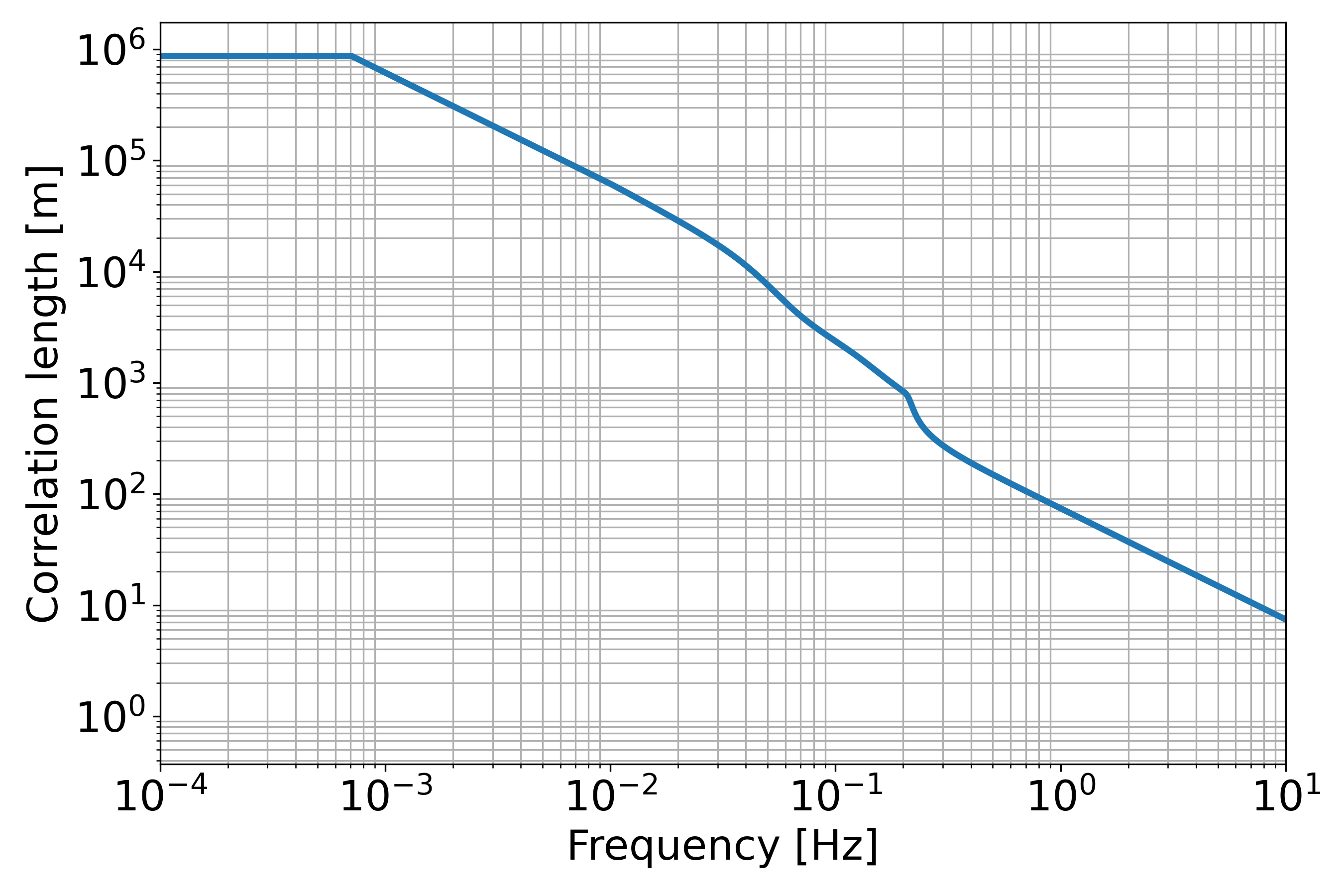}
\includegraphics[width=0.48\textwidth]{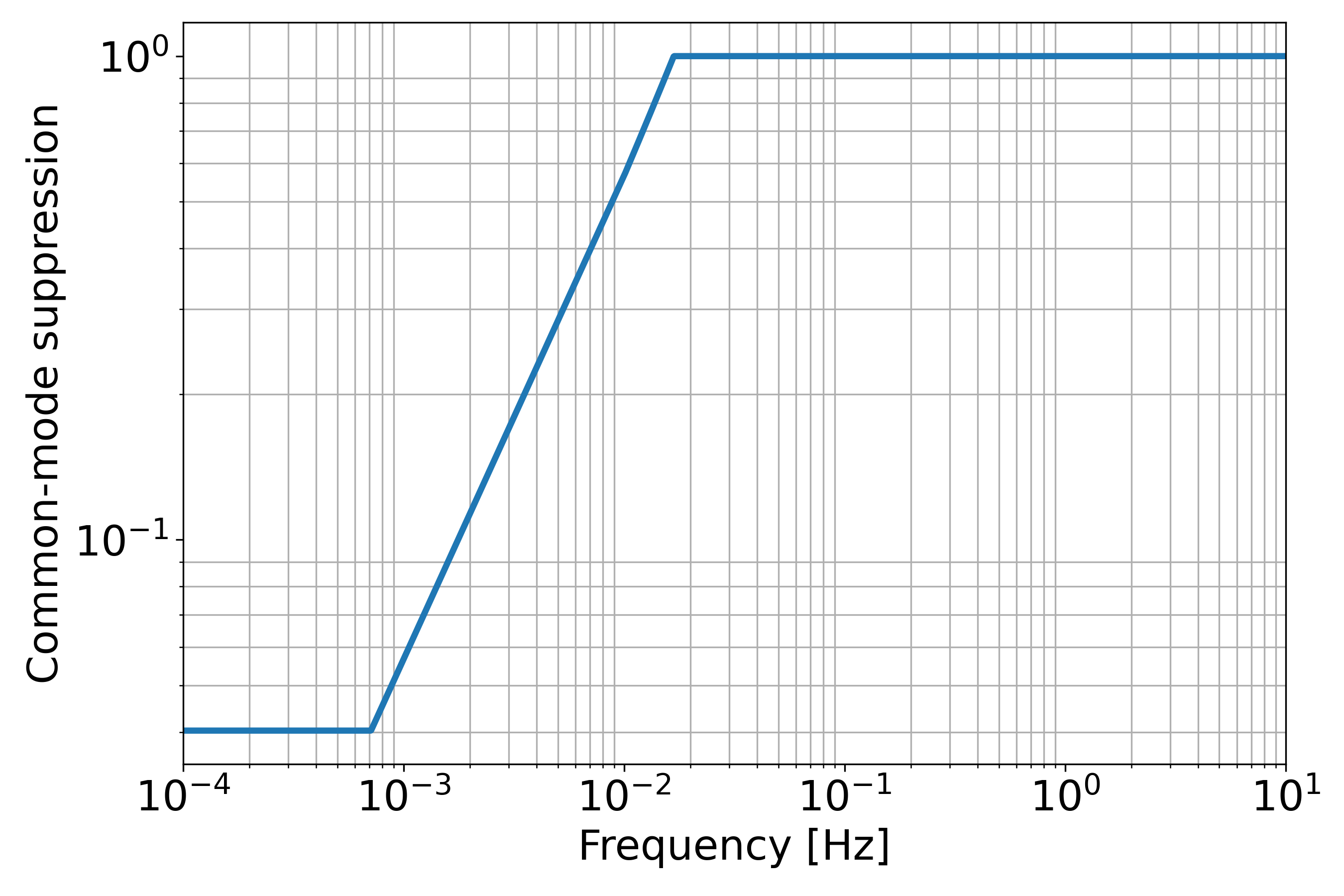}}
\caption{Left: Plot of the correlation length of a fundamental Rayleigh-wave field estimated from seismic speeds using a layer model obtained from Garcia et al \cite{GaEA2019}. Seismic scattering, mode content and source distribution are not considered here, but might have an important influence on correlations. Right: Common-mode suppression of seismic displacement measured across a distance of $L=35\,$km calculated from the correlation model.}
\label{fig:seiscorr}
\end{figure}
Another important aspect are seismic correlations between points on the lunar surface. They determine the LGWA array configuration, and they also influence the seismic noise of LBI-GND concepts, which measures ground strain. Strong correlations (with matching phase) between the ends of an LBI-GND detector means that it sees a smaller seismic strain, which reduces the impact of the seismic background on the GW measurement. We call this effect \emph{common-mode rejection}. Correlations are determined by seismic speeds, seismic scattering, source distribution and mode content of the seismic field. In figure \ref{fig:seiscorr}, we show an estimate of the correlation length obtained from a simulated Rayleigh-wave dispersion curve. The layered geological model is taken from Garcia et al \cite{GaEA2019} and using the gpdc tool of the Geopsy pack to calculate a Rayleigh-wave dispersion curve \footnote{\url{https://www.geopsy.org/index.html}}. Assuming a LBI-GND arm length of 35\,km, we find that above 20\,mHz, ground displacement is uncorrelated. The lower bound on the common-mode rejection is set by the longest spatial scales of the displacement field. In theory, there are modes without gradient along great circles (all $l=0$ modes), but we choose a more conservative value of $R_{\rm moon}/2$ since the actual correlation also depends on which modes dominate the hum around a mHz. One should also expect stronger seismic gradients because of topography and heterogeneous geology.

\subsection{Seismic isolation}
Seismic isolation is the defining feature of the LBI-SUS concept. Seismic isolation generally consists of active \cite{MaEA2015} and passive \cite{AcEA2010} elements. Active seismic isolation is a form of vibration control using sensors to stabilize a platform. A chain of spring or pendulum stages can then be suspended from this platform to filter out disturbances above its fundamental resonances, by which we mean the system of coupled resonances forming from the fundamental resonances of the individual stages. The final suspension stage holds the test mass. All forms of ground motion (tilt and displacement in any direction) can couple into motion of the test mass along the direction of the interferometer arm leading to seismic noise of the GW measurement. 

\begin{figure}
\centering{
\includegraphics[width=0.8\textwidth]{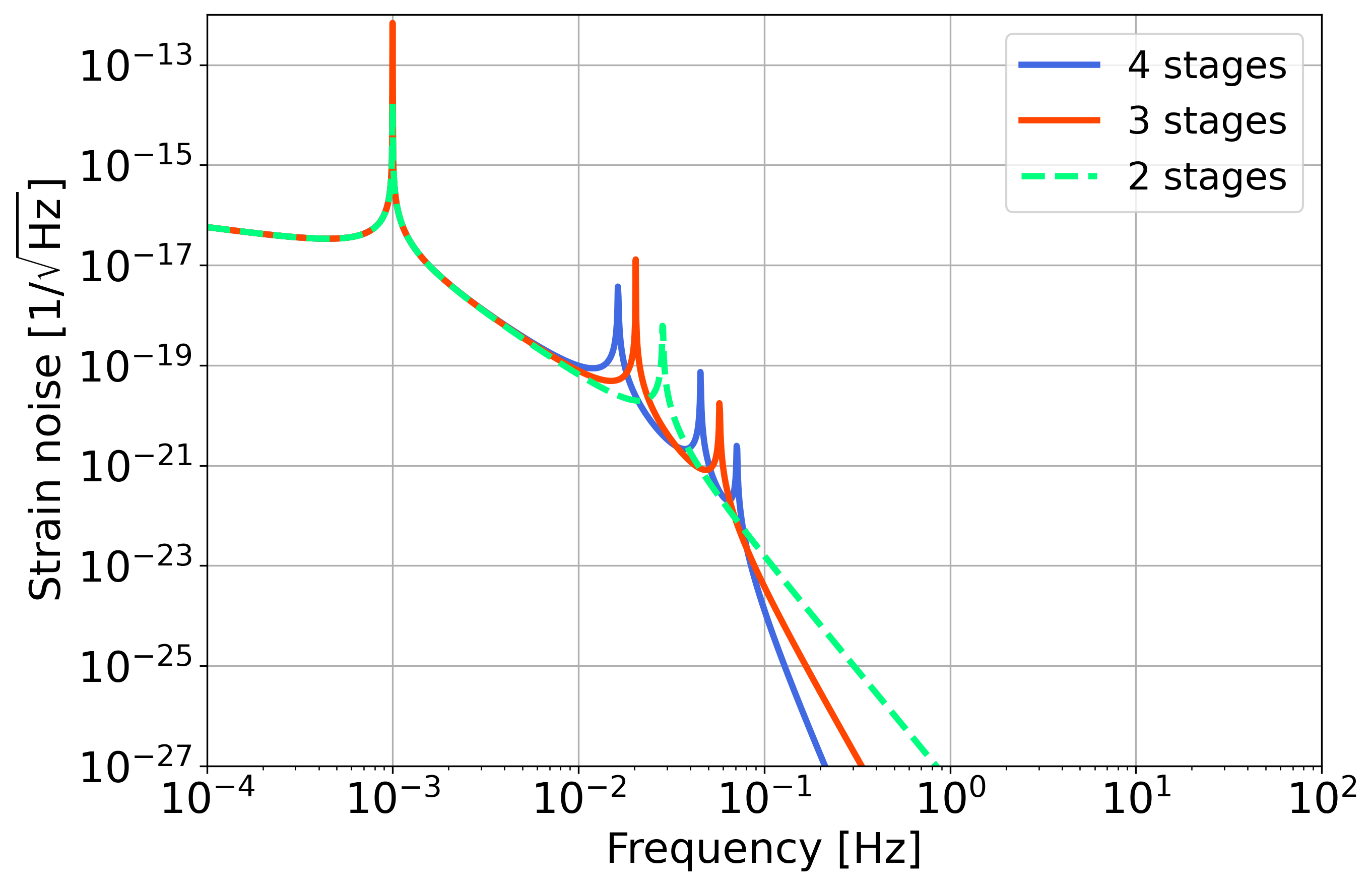}}
\caption{Comparison of strain noises from horizontal ground displacement. The suspension systems consist of $n-1$ 20\,mHz stages and a final 1\,mHz stage. Here we assume uncorrelated ground motion between the different LBI-SUS test masses, equation (\ref{eq:seisspectrum}) for the PSD of ground displacement, and 35\,km long interferometer arms. The suspended masses are 800\,kg (top mass of 4-stage system), 800\,kg (top mass of 3-stage system), 1200\,kg (top mass of 2-stage system), 1289\,kg (test mass). }
\label{fig:seisisol}
\end{figure}

Here we will focus on the modeling of seismic isolation with respect to ground displacement along the horizontal direction of the interferometer arm and along the vertical. We model the suspension system as a sequence of suspended masses $m_k,\,k=1,\ldots,n$, where $n$ is the number of suspended stages and $m_n$ is the test mass. Each suspension stage, e.g., a pendulum for horizontal filtering or a system of spring blades for vertical filtering, is characterized by a resonance frequency $\omega_k$. We also define $m_{ij}=m_i+\ldots+m_j$ with $i<j$, e.g., $m_{13}=m_1+m_2+m_3$, and $m_{ii}=m_i$. The equation of motion for mass $m_k$ can then be cast into the form
\begin{equation} 
 - m_k \omega^2 \tilde x_k  +  m_{kn}\left(\tilde x_k - \tilde x_{k-1}\right) (1 + i\phi_k)\omega_k^2+  m_{(k+1)n}\left(\tilde x_k - \tilde x_{k+1}\right) (1 + i\phi_{k+1})\omega_{k+1}^2 = \delta \tilde F_k,
 \label{eq:susmech}
\end{equation}
where $\phi_k$ is the loss angle of a suspension stage. A loss angle quantifies the fraction of energy in the system lost due to dissipation. When an index in this equation takes a value smaller than 1 or larger than $n$, the respective variable must be set to 0. Instead of considering a force $\delta \tilde F_k$ acting on the mass $m_k$, it can sometimes be more convenient to model the external disturbance as displacement noise. This is the case for the seismic noise coupling into the system through the top most suspension point. In this case, the equivalent force acting on the top mass $m_1$ takes the form $\delta\tilde F_1=m_{1n}\omega_1^2\tilde x_0$, where $\tilde x_0$ is the displacement of the suspension point of the first stage. A comparison of the isolation performance of suspension systems with a different number of stages is shown in figure \ref{fig:seisisol}.

The equations of motion neglect coupling between displacements along orthogonal directions, which is a good approximation for the purpose of this paper. However, even if the mechanical system is engineered so accurately that cross couplings between orthogonal displacements can be ignored, it turns out that vertical displacement of the test mass couples into the GW measurement. This is because the vertical direction is defined by the direction of local gravity. This direction is different at the two ends of an interferometer arm mostly due to the curvature of the Moon. If the interferometer has an arm length $L$, the vertical to horizontal coupling is given by $L/(2R)$, where $R$ is the radius of the Moon. For a 35\,km long arm, we find a 1\% vertical to horizontal coupling. This has important consequences for the seismic isolation system as we explain in the following.

It is unfeasible to achieve the 20\,mHz and 1\,mHz resonances with conventional pendulum stages or spring assemblies, which would all result in much higher resonance frequencies. Instead, a mechanical system would rely on spring-antispring dynamics \cite{Win2002}. Here, antispring dynamics means that the elastic restoring force has the opposite sign compared to a normal spring as can for example be found in an inverted pendulum when the gravitational pull on the supported mass exceeds the elastic restoring force of the supporting bar \cite{SaEA1994}. While it is conceivable that 20\,mHz and maybe even 1\,mHz resonances can be achieved for horizontal filters \cite{WLB1999}, it is unthinkable today to achieve such values for the vertical isolation. This is because the vertical stages must also counteract the gravitational pull of the Moon, i.e., the challenge is to realize a soft suspension dynamics on top of a strong force that counteracts gravity. Euler springs, which are assemblies of near vertically aligned cantilevers, which provide a spring-like restoring force through lateral deformation (buckling), were identified as promising mechanical stages for vertical isolation, but with performance far from what is needed for an LBI-SUS \cite{vHEA2023a}. There is an additional challenge concerning the suspension thermal noise of the lower suspension stages, which points to even more exotic suspension systems like superconducting magnetic levitation. In any case, we believe that vertical seismic isolation is a potential show-stopper for LBI-SUS concepts. 

In a real suspension system, the resonances would be damped using feedback control designed for this purpose, but since it does not play a role for our sensitivity studies, we use the undamped suspension model.

\subsection{Newtonian noise}
The lunar seismic field also produces fluctuations of the gravitational field. The corresponding noise in GW measurements is called Newtonian noise (NN). It only affects LBI-SUS concepts. In this study, we use a very simple model that only considers NN contributions from the displacement of the lunar surface. Contributions from the compression of the ground are neglected. We also assume for simplicity that the ambient seismic field is isotropic and homogeneous. In this case, the NN is given by \cite{Har2019}
\beq
h_{\rm NN}(f)=\frac{2}{L}\frac{2\pi G\rho_0}{\sqrt{2}}\frac{\xi(f)}{(2\pi f)^2},
\eeq
where $\xi(f)$ is the vertical surface displacement, $G$ Newton's constant, and $\rho_0$ the density of the ground. We will see in section \ref{sec:ifo} that NN is too weak to play an important role for lunar GW detection. However, there can be stronger NN transients produced by moonquakes. In this case, a cancellation of NN can be attempted, which requires the deployment of seismometer arrays around all test masses \cite{Cel2000,BaEA2020}. 

\subsection{Suspension thermal noise}
The coupling between the mechanical system and a heat bath of temperature T leads to thermal fluctuations described by the \textit{fluctuation-dissipation} theorem. Accordingly, the suspension thermal noise spectrum is determined by the real part of the complex admittance $Y(\omega) = -\irm\omega \tilde{x}(\omega)/\tilde{F}(\omega)$, with $\tilde{x}(\omega)$ being the displacement amplitude of a suspended mass at frequency $\omega$ and $\tilde{F}(\omega)$ the force acting on it \cite{HaML2018}. One can think of the real part of the admittance of a mechanical system as the ease, with which it reacts to a probing force. A high-admittance system will experience larger displacements of its components than a low-admittance system under the same force. Given this, the PSD of the fluctuations can be written as:
\begin{equation}
    S(x;\omega) = \frac{4\mathrm{k_B}T}{\omega^2}\Re(Y(\omega)).
    \label{eq:Levin}
\end{equation}
To understand this results, let us take a simple pendulum as an example; we introduce a term of structural damping, parameterized by the so-called \textit{loss angle} $\phi$, which determines a small imaginary part of a restoring force. We can write the equation of motion of this pendulum of length $L$ with suspended mass $m$ in the frequency domain:
\begin{equation}
    \omega_0^2\Big(1+\frac{L_{\rm el}}{L}(1-\irm\phi/2)\Big)\tilde{x}(\omega) - \omega^2\tilde{x}(\omega) = \delta\tilde{F}(\omega)/m
\end{equation}
with $\omega_0^2 = g/L$; the parameter $L_{\rm el}$ is the bending length of the suspension fiber \cite{CaEA2000}, and $L_{\rm el}/L$ is the fraction of elastic to gravitational restoring force of the pendulum also known as the dilution factor. This dilution factor enhances the quality factor of the pendulum mode with respect to the material quality factor. It does not exist for filter stages of vertical displacement, where the restoring force cannot get any assistance from gravity. The admittance is found to be:
\begin{equation}
    \tilde{Y}(\omega) = -\irm\omega \frac{\tilde{x}}{\delta\tilde{F}} = \frac{-\irm\omega}{m(\omega_0^2-\omega^2+\omega_0^2(L_{\rm el}/L)(1-\irm\phi/2))}
\label{eq:suscept}
\end{equation}
Calculating its real part and inserting the result in equation (\ref{eq:Levin}), we obtain the following PSD of the thermal fluctuations of the mass coordinate $x$:
\begin{equation}
    S(x;\omega) = \frac{4\mathrm{k_B}T}{m\omega}\frac{\omega_{\rm el}^2\phi/2}{(\omega_0^2+\omega_{\rm el}^2-\omega^2)^2+\omega_{el}^2\phi^2/4}
\label{eq:psd_susther2}
\end{equation}
with $\omega_{\rm el}^2=\omega_0^2(L_{\rm el}/L)\ll \omega_0^2$. Note that the quality factor effectively decreases when the resonant frequencies are reduced by spring-antispring dynamics \cite{SaEA1994}, which means that setting a requirement for a Q-factor at low resonance frequency poses a greater challenge to meet this requirement than if the same Q-factor had to be achieved with a higher resonance frequency \cite{HaML2018}. 

For multi-stage suspension systems, the susceptibility in equation (\ref{eq:suscept}) can be calculated by solving the system of equations (\ref{eq:susmech}) applying a force $\delta\tilde{F}_n$ to the test mass $m_n$. The dominant thermal noise is produced by the final suspension stage. Thermal noise from upper stages are filtered out to some extent by the suspension system, but can still be important if the quality factor of the upper stages is substantially lower than the final stage. Moreover, the vertical-to-horizontal coupling, which we already discussed in the context of seismic isolation, also plays an important role for suspension thermal noise of LBI-SUS concepts. Since 1\% of vertical motion couples into the GW measurement (assuming a 35\,km long detector), it means that the quality factor of vertical isolation stages cannot be less than 4 orders of magnitude smaller than the diluted quality factor of horizontal isolation stages. Otherwise, the vertical stages would dominate the suspension thermal noise.

\begin{figure}
\centering{\includegraphics[width=0.8\textwidth]{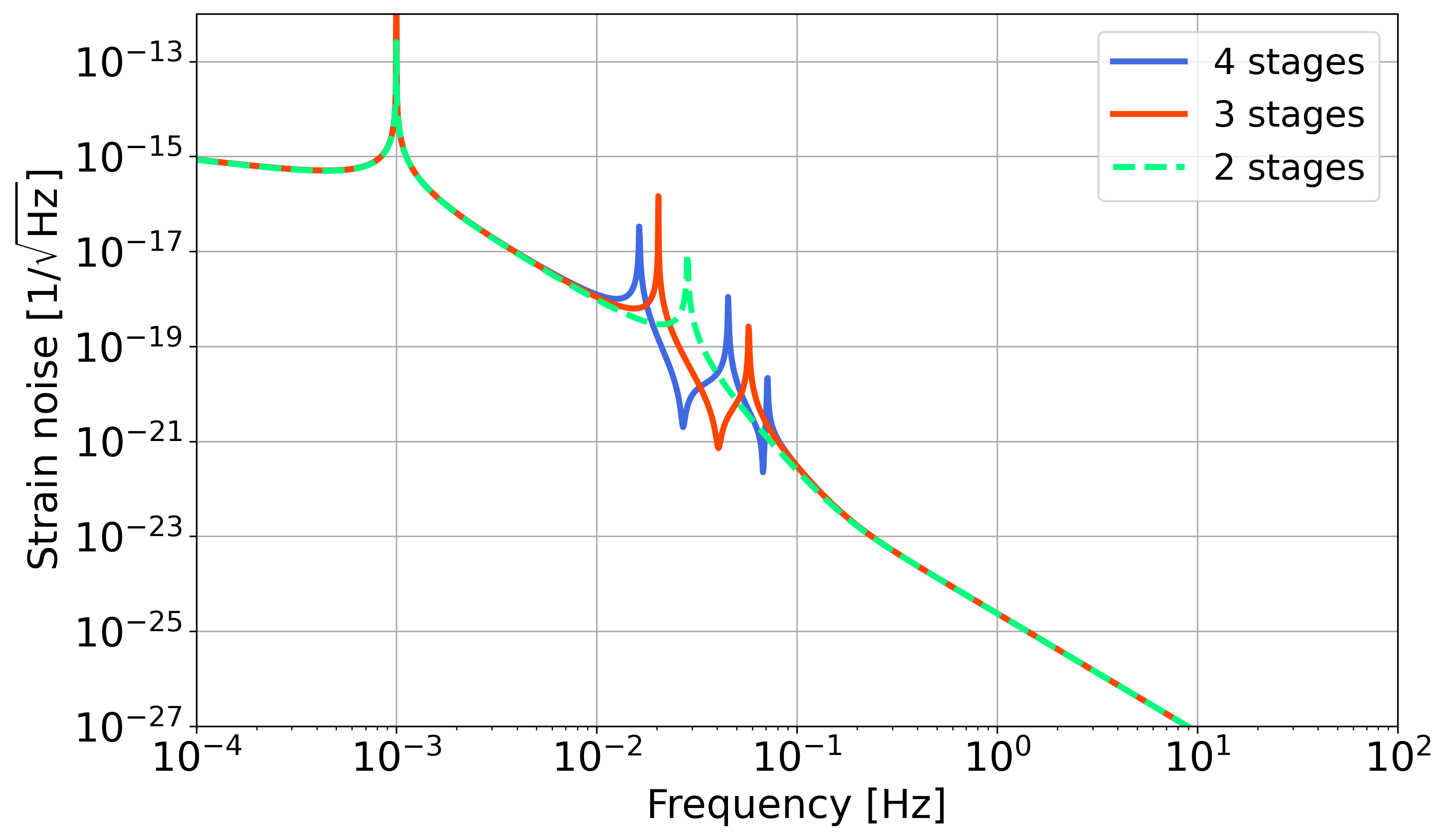}}
\caption{Suspension thermal noise of systems with two (dashed green), three (red), and four (blue) stages of attenuation. The penultimate stage makes the main contribution to the noise up to 0.1\,Hz, while the noise from the final stage dominates above 0.2\,Hz.}
\label{fig:confrontth}
\end{figure}

We have computed the suspension thermal noise in the case of two, three and four suspension stages, and we show their trends in figure \ref{fig:confrontth}. The final stage has a 1\,mHz resonance while all other stages have a 20\,mHz resonance. We assumed a loss angle of $10^{-9}$ for the final stage, and $10^{-3}$ for all other stages. As we can see from this plot, the strain noise is smaller than $10^{-23} \,\mathrm{Hz}^{-1/2}$ at $\sim 200$\,mHz in all the three cases. 
These frequencies and loss angles in this example were chosen to meet the requirements for decihertz GW detection. However, there is no known or envisioned technology that would enable such low loss factors combined with the very low resonance frequencies. Furthermore, as pointed out before, assuming similar resonance frequencies for the vertical suspension stages, the final stage would need a loss factor no larger than $10^{-5}$. 

\subsection{Mirror thermal noise}\label{sec:coating}
Mirror thermal noise is connected to dissipation mechanisms in the mirror substrate and coating layers. Mechanical dissipation leads to Brownian noise, which is the dominant contribution to mirror thermal noise in current GW detectors \cite{HaEA2002}. Thermal fluctuations can lead to thermoelastic noise and thermorefractive noise \cite{EvEA2008}. Coatings are formed by multiple layers of materials with a typical optical thickness of $\lambda/4$. Each coating layer is obtained from a deposition of atomic layers over the substrate's surface, e.g., through ion-beam sputtering. A full coating stack has a few tens of layers and each layer typically has sub-micron thickness. The noise contribution of this inhomogeneous multi-layer coating can be calculated using Levin’s method \cite{Lev1998a} of calculating thermal noise from the fluctuation-dissipation theorem written in equation (\ref{eq:Levin}). The Levin approach does not give an answer though to how exactly the losses in the substrate and the multi-layer coating contribute to the overall loss; it was later provided through detailed studies of the heterogeneous structure \cite{HaEA2002, HoEA2013}. Since our coating-thermal noise estimate does not need to be very accurate, we choose a simplified model that only takes into account mechanical loss of substrate and coating, and we use the Nakagawa approximation \cite{NaEA2002},
\begin{equation}
    S^{\textrm{MTN}}_x(f) = \frac{2\mathrm{k_B}T}{\pi^{3/2} f}\frac{1-\sigma^2}{Y w} \Big(\phi_{\textrm{sub}} + \frac{2}{\pi^{1/2}} \frac{d}{w}\frac{1-2\sigma}{1-\sigma}\phi_{\textrm{coat}} \Big),
\label{eq:CoatTh}
\end{equation}
where we have introduced the Young's modulus $Y$ and Poisson's ratio $\sigma$ of the substrate, $d$ is the thickness of the multi-layer coating, $w$ is the beam radius on the mirror, and $\phi_{\textrm{sub}},\,\phi_{\rm coat}$ are the substrate and coating loss angles. 
\\
If the interferometer has arm cavities of length $L$ and with curvature radii $R_1,\,R_2$ of the input and end test masses, we can calculate the beam radius using
\begin{equation}
    w(z) = w_0\sqrt{1+\Big(\frac{z}{z_R}\Big)^2},
\label{eq:beamrad}
\end{equation}
where $z_{\rm R} = \pi w_0^2/\lambda_0$ is the Rayleigh range, and $z$ is the distance from the waist of the laser beam along the beam axis, where the waist is the minimal diameter of a beam transverse to the optical axis. The waist is located at a distance $z_0 = Lg_2(1-g_1)/(g_1+g_2-2g_1g_2)$ from the input test mass, where $g_1 = 1-L/R_1$ and $g_2 = 1-L/R_2$ are the g-factors. This leads to a minimal beam radius of  $w_0 = (\lambda_0L/\pi)^{1/2}(g_1g_2(1-g_1g_2)/(g_1+g_2-2g_1g_2)^2)^{1/4}$. We can then find the beam radii on the two test masses by calculating $w(z_0)$ and $w(L-z_0)$.

The mirror design of an LBI detector inherits some of the known issues of Earth-based detectors working at cryogenic temperatures \cite{AdEA2020}. Materials need to be compatible with the cryogenic temperatures in a lunar PSR ($T = 40 - 100\, \unit{\kelvin}$). This influences the choice of laser frequency and design of the coating stack. Not only are we interested in the loss angle, but also in achieving low absorption and scattering. Apart from the material properties, we also need to highlight that the size of the test masses needs to be very large ($\mathcal{O}(\mathrm{ton})$), which begs the question on whether it is possible to deposit the required high-quality coatings on such large substrates.

\subsection{Dust noise}
\label{sec:dust}
Given the strict requirements on residual gas and dust contamination in current detectors, noise from lunar dust was pointed out as a potential show-stopper for LBI-concepts on the Moon. Dust and micro-meteoroids have an extremely low density in space, and in fact, a study showed that noise introduced by inter-planetary particles can be neglected for the LISA detector \cite{RPH2009}. However, meteoroids act as impactors on the lunar surface, which generates ejecta particles from the lunar regolith. There is a continuous bombardment of the lunar surface by meteoroids, which leads to a continuously present particle cloud around the Moon. The dust distribution around the Moon was measured with the Lunar Dust Experiment (LDEX) on the Lunar Atmosphere and Dust Environment Explorer (LADEE) mission for grain sizes larger 0.3\,$\mu$m \cite{HoEA2015}. The measurement was done at altitudes between a few kilometers and 260\,km, and spanning selenographic latitudes of -23$\degree$ to +24$\degree$. The observed particle concentration was highest versus the apex direction (direction of motion of the Earth-Moon system around the Sun) consistent with impactor sources in the equatorial plane. At lower altitudes, the observed particle density varied around a few $10^{-3}$\,m$^{-3}$. A model consistent with these observations was later developed to extend the distributions to the polar regions and lower altitudes \cite{SzEA2019}. Their model predicts particle densities around $10^{-3}$\,m$^{-3}$ at the poles. They estimate an average speed of impact ejecta around 660\,m/s. Since the distributions are matched to the LDEX measurements, which did not reach altitudes below 1\,km, this model might miss a near-surface, low-speed ejecta population.

The mechanism by which dust particles produce phase noise is through forward scattering. The strain noise is given by \cite{RPH2009}
\beq
h_{\rm DN}(f)=|\Re(2(\tilde m-1))|\sqrt{\frac{1}{Lw_0\bar v n}}\exp\left(-\sqrt{2\pi\lambda L}f/\bar v\right),
\label{eq:scatter}
\eeq
where $\Re(\cdot)$ is the real part, $n$ is the particle number density, $\bar v$ the average particle speed, and $\tilde m$ is connected to the forward scattering amplitude $S(0)$ of the dust particle,
\beq
\tilde m=1+\irm \frac{2\pi n}{k^3}S(0),
\eeq
and $k$ is the wavenumber of the laser beam.

\begin{figure}
\centering
\includegraphics[width=0.7\textwidth]{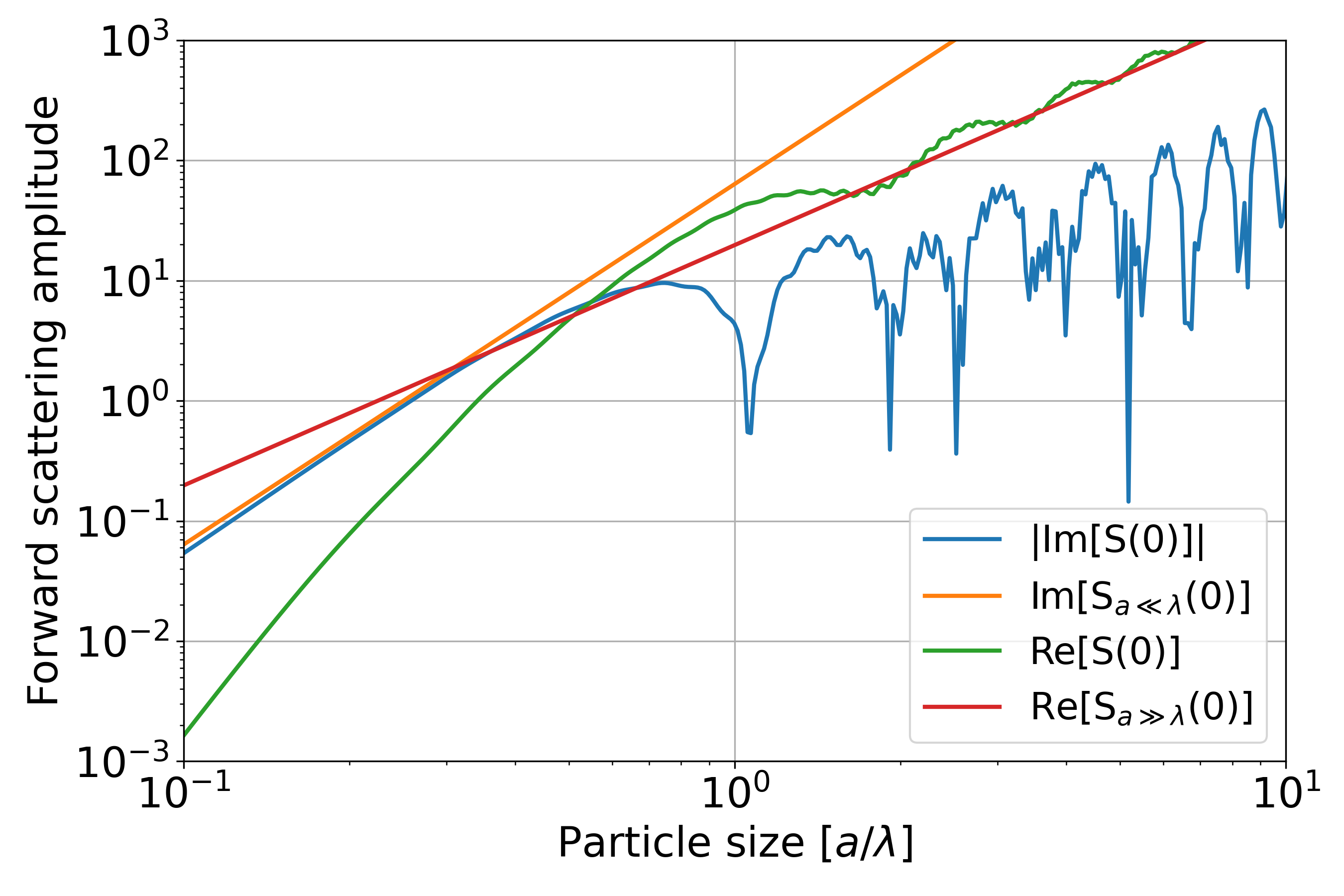}
\caption{Forward scattering amplitudes for Mie scattering. The orange line is the small-particle approximation (Rayleigh scattering) of equation (\ref{eq:rayleigh}). The red line is the large-particle approximation of equation (\ref{eq:largepart}). The blue and green curves are the real (absorption) and imaginary (refraction) part of the Mie scattering amplitude including terms up to $n=100$ in equation (\ref{eq:miescatt}). The imaginary part determines the phase noise according to equation (\ref{eq:scatter}).}
\label{fig:miescattering}
\end{figure}

Since the typical diameter $a$ of dust particles is similar to the laser wavelength $\lambda$, we are in the Mie-scattering regime. Before presenting an estimate of Mie-scattering amplitudes $S(0)$, we briefly discuss the two simpler cases where the particles are much larger or much smaller than $\lambda$. For homogeneous, non-absorbing, spherical particles with $a\ll\lambda$, we obtain the equation for Rayleigh forward scattering \cite{Hul1981}
\beq
S(0)=\irm k^3\alpha=\irm \frac{1}{3}(m^2-1)(ka)^3
\label{eq:rayleigh}
\eeq
where $\alpha$ is the polarizability, and $m$ is the refractive index of the particle. In general, the polarizability can have an imaginary part describing absorption. In the case of $a\gg\lambda$ and if the particle is spherical, we have
\beq
S(0)=\frac{1}{2}(ka)^2.
\label{eq:largepart}
\eeq
Since this scattering amplitude is real-valued, it means that it describes pure absorption. Concerning phase noise produced by dust, the most interesting regime is Mie scattering. For spherical particles, we can expand the forward scattering amplitude into a sum according to \cite{Hul1981}
\beq
S(0)=\frac{1}{2}\sum\limits_{n=1}^\infty(2n+1)(a_n+b_n),
\label{eq:miescatt}
\eeq
where the expansion coefficients are given by
\beq
\begin{split}
a_n &= \frac{\psi^\prime(y)\psi_n(x)-m\psi_n(y)\psi_n^\prime(x)}{\psi^\prime(y)\zeta_n(x)-m\psi_n(y)\zeta_n^\prime(x)},\\
b_n &= \frac{m\psi^\prime(y)\psi_n(x)-\psi_n(y)\psi_n^\prime(x)}{m\psi^\prime(y)\zeta_n(x)-\psi_n(y)\zeta_n^\prime(x)}.
\end{split}
\eeq
These expressions depend on the spherical Bessel function and on the spherical Hankel function of the second kind:
\beq
\psi_n(z)=zj_n(z),\,\zeta_n(z)=zh^{(2)}_n(z),
\eeq
where $y=mka,\,x=ka$. We use this expression to evaluate the phase noise in LBI concepts. Figure \ref{fig:miescattering} shows the scattering amplitudes for absorption and refraction.

\section{Long-baseline, laser-interferometric GW measurement}
\label{sec:ifo}

\subsection{LBI-GND}
\begin{figure}
\centering
\includegraphics[width=0.8\textwidth]{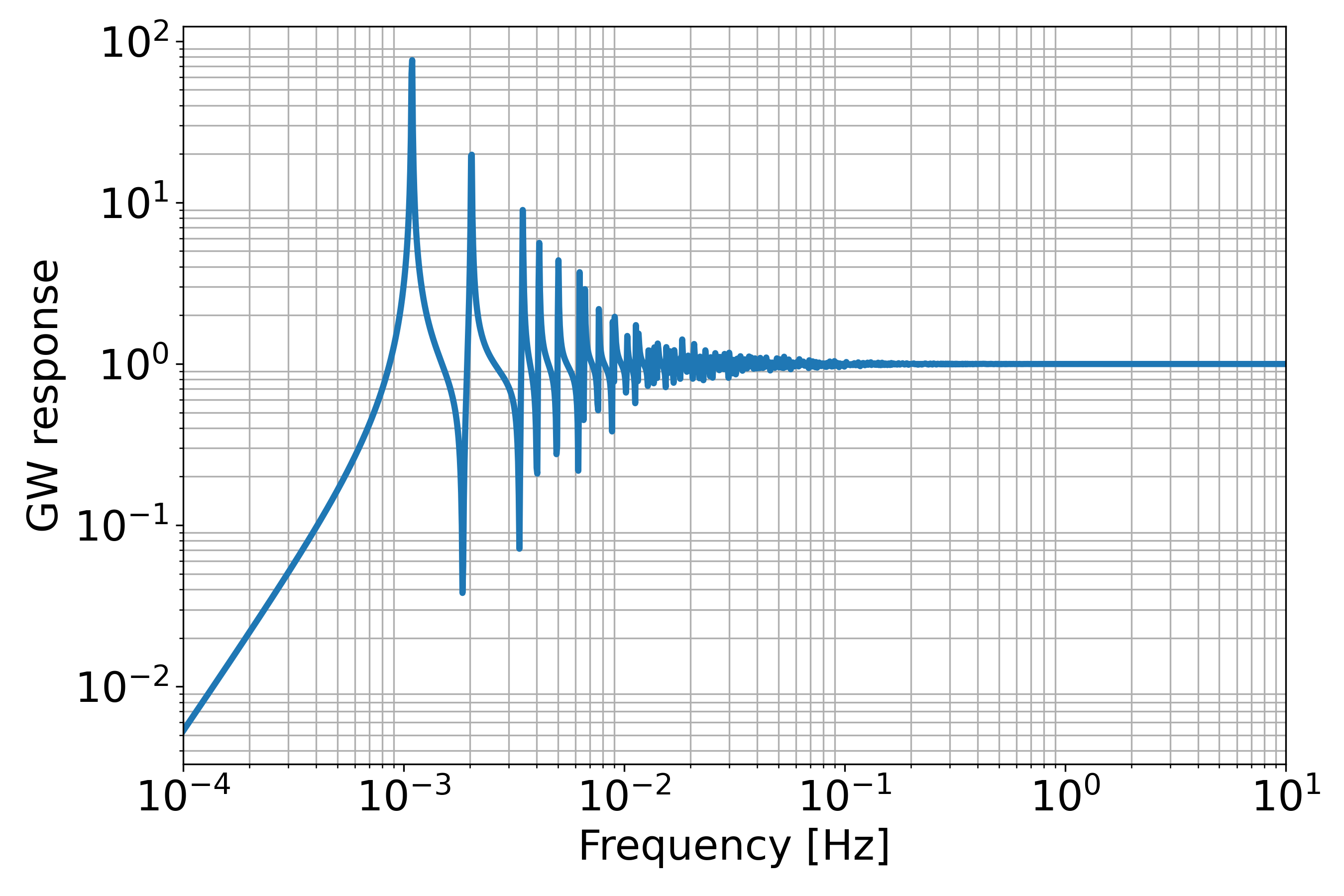}
\caption{Lunar GW response in terms of seismic strain along the surface used for the LBI-GND concept. The Moon's response to GWs does not affect the LBI-SUS signal due to its seismic-isolation system.}
\label{fig:LSGAresp}
\end{figure}

The LBI-GND concept is a long-baseline laser interferometer to measure ground strain. This strain is produced by GWs as well as by the seismic background. A model of the elastic strain response of the Moon to GW strain is shown in figure \ref{fig:LSGAresp}. 
The Moon behaves like a fluid above a few 10\,mHz, which means that its mass distribution follows the GW quadrupole field without significant elastic resistance like an assembly of freely falling test masses. At lower frequencies, the response shows amplification on the resonances of the quadrupole normal modes. In this model, the low-frequency modes are assumed to have quality factors of about 100. Below the lowest order quadrupole mode at about 1\,mHz, the GW response of the Moon is strongly suppressed by its stiffness. This response function is used to convert instrument noises of the seismic strain measurement into equivalent noises of the GW measurement.

\begin{figure}
\centering
\includegraphics[width=0.8\textwidth]{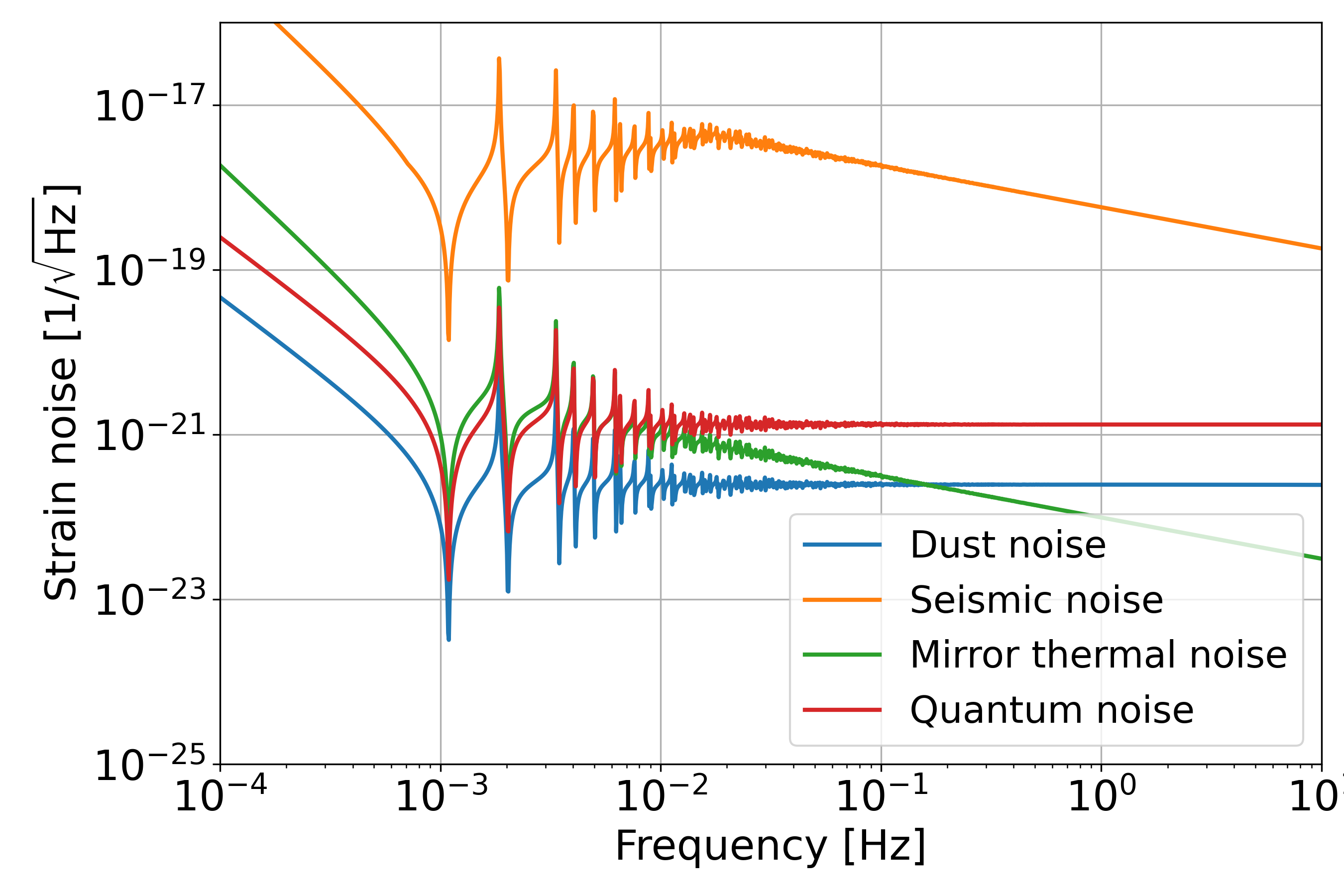}
\caption{Some of the principal noises of a long-baseline laser-interferometric lunar GW detector with optics mounted on the ground. The DN model assumes a particle density of $1\,\unit{\kilo\meter}^{-3}$ near the lunar surface.}
\label{fig:LSGAsens}
\end{figure}

The LBI-GND concept does not require seismic isolation systems and it does not require suspended test masses. The optics are directly mounted to the ground. This also means that suspension thermal noise and radiation-pressure noise do not play a role. The noises we include in the model are quantum noise, mirror thermal noise, dust noise, and the noise produced by the seismic background. The parameter values of these models are summarized in table \ref{tab:lbigndtab}. As for LGWA, we assume that the LBI-GND detector would be deployed inside a PSR to avoid disturbances due to strong temperature variations. More precisely, the three stations would need to be deployed on the crater wall of a PSR to have a clear line of sight between the interferometer stations. In this case, the ambient temperature is below 100\,K. There are substrate and coating materials with low mechanical loss at low temperatures such as silicon \cite{NaEA2008}, but mirror thermal noise is a very small contribution to the instrument noise for LBI-GND detectors. We choose fused silica with silica-tantala coatings in our model taking into account the increased mechanical loss at low temperature \cite{Sch2008}. 
\begin{table}[h!]
    \centering
    \begin{tabular}{lc}
    \hline
        Parameter  & Value \\
        \hline
        \hline
        \textbf{Relevant geophysical parameters}&\\
        \hline
        Temperature & 100\,\unit{\kelvin}\\
        Speed of dust particles & 660\,\unit{\meter /\second}\\
        Dust particle density& $10^{-3}$\,\unit{\meter^{-3}}\\
        Refractive index of dust particle& 1.33\\
        \hline
        \textbf{Detector parameters} & \\
        \hline
        Arm length & 35\,\unit{\kilo \meter} \\
        Radius of end mirrors & 0.4\,m \\
        Laser input power& 10\,W\\
        Laser wavelength & 1064\,nm\\
        Test mass material & Fused silica\\
        Coating loss angle &  $10^{-3}$\\
        Substrate loss angle &  $10^{-4}$\\
        \hline
        \hline
    \end{tabular}
    \caption{Summary of the model parameters of the LBI-GND configuration.}
    \label{tab:lbigndtab}
\end{table}
The main instrument parameters were chosen so that it should not be too hard to achieve them, but at the same time realizing a readout sensitivity such that the seismic background becomes the clearly dominant noise. As shown in figure \ref{fig:LSGAsens}, seismic noise is stronger than other instrument noise by a few orders of magnitude with our model. 

While this LBI-GND concept seems feasible on paper, it must be emphasized that it is still a complex system whose deployment would face substantial challenges. Three stations would have to be deployed to form a full Michelson interferometer. This is necessary to reduce the laser frequency noise in the GW measurement. The optics must be properly aligned to be able to form the interferometer. Maybe it is possible to remove laser frequency noise like in LISA through electronic combinations of signals with appropriate time delays \cite{Tin2014}. However, the much shorter arm length and higher laser power of the LBI-GND concept compared to LISA would require significant modifications. It is beyond the scope of this paper to analyze implementation details, but we do not see clear show-stoppers for the LBI-GND concept. We also point out that optical fibers were proposed to create the long-baseline interferometric readout, but this would come at a loss of sensitivity, e.g., due to fundamental thermal noise of the fiber and scattering \cite{HiEA2022}. A fiber interferometer is not suitable for GW measurements.

\subsection{LBI-SUS}
The LBI-SUS concept works similar to LIGO-Virgo interferometers. It requires seismic isolation and suspension systems. Certainly, it is by far the most challenging concept to realize, and in fact, given our current experience with terrestrial GW detectors, it would be impossible to build and operate them on the Moon. Revolutionary advances in engineering would be required to deploy and commission the detectors without continuous intervention by humans. Nonetheless, here our attention is not on the engineering, but on the most basic aspects of the detector design.
\begin{table}[h!]
    \centering
    \begin{tabular}{lc}
    \hline
        Parameter  & Value \\
        \hline
        \hline
        \textbf{Relevant geophysical parameters}&\\
        \hline
        Temperature & 100\,\unit{\kelvin}\\
        Speed of dust particles & 660\,\unit{\meter /\second}\\
        Dust particle density& $10^{-3}$\,\unit{\meter^{-3}}\\
        Refractive index of dust particle& 1.33\\
        \hline
        \textbf{Detector parameters} & \\
        \hline
        Arm-cavity length & 35\,\unit{\kilo \meter} \\
        Radius of TM & 0.4\,\unit{\meter} \\
        Thickness of TM & 1.1\,\unit{\meter} \\
        Chain of suspended masses & 800/1200/1289\,\unit{\kilo \gram} \\
        Loss angle of final stage & $10^{-9}$\\
        Loss angles of upper two stages & $10^{-3}$\\
        Resonance frequency final stage & 1\,m\unit{\hertz}\\
        Resonance frequencies of upper two stages & 20\,m\unit{\hertz} \\
        Amplitude transmissivity of ITM & 0.1\\
        Laser arm-input/arm-cavity power & 0.1\,\unit{\watt}/40\,\unit{\watt}\\
        Laser wavelength & 1500\,\unit{\nano\meter}\\
        Test mass material & Silicon\\
        Coating loss angle &  $10^{-5}$\\
        Substrate loss angle &  $10^{-8}$\\
        ITM radius of curvature & 17.7\,\unit{\kilo \meter}\\
        ETM radius of curvature & 17.7\,\unit{\kilo \meter} \\
        TM pitch resonance frequency & 1\,\unit{\milli\hertz}\\   
        \hline
        \hline
    \end{tabular}
    \caption{Overview of the LBI-SUS model parameters. The masses of the suspension system are listed from top mass to test mass.}
    \label{tab:lbisustab}
\end{table}

We consider an LBI-SUS model with 35\,km arm length. We assume that arm cavities are required. They add complexity to the system, but they provide an important handle to control and reduce noise couplings from auxiliary degrees of freedom, which might prove crucial for the ambitious LBI-SUS concept. Each test mass is suspended from a chain of passive isolation stages, e.g., pendula and spring blades for horizontal and vertical isolation. We consider a suspension system with 3 stages. The final stage cannot be mechanical since it would be impossible to meet the requirements on mechanical loss and resonance frequency. As a possible solution, we propose magnetic levitation with superconducting coils. The idea was already investigated for terrestrial GW detectors \cite{varvella2004}. It might be equally impossible to realize this stage through levitation, but it cannot be ruled out as a possibility given our current level of understanding of superconducting magnetic levitation. None of the stages can be realized as simple springs or pendula since it would require enormous mechanical structures to achieve such low resonance frequencies. Mechanical spring-antispring stages might be an option for all but the final stage \cite{WLB1999,vHEA2023a}.

As for the other lunar GW detector concepts, we consider deployment in a PSR for improved temperature stability, and the deployment of the three interferometer stations would have to be on the crater walls to have a clear line of sight between the stations. The Moon's cryogenic temperatures set constraints on the choice of the test-mass substrate and coating materials. For our model, we choose silicon for the test mass, which has low mechanical loss at cryogenic temperatures \cite{NaEA2008}. This is a common choice also in the case of third generation interferometers \cite{AdEA2020,ET2020}. The bulk loss of silicon is very low and therefore, the bulk contribution to the mirror thermal noise is negligible with respect to typical loss angles for the coating materials; see equation (\ref{eq:CoatTh}).

\begin{figure}[h!]
\centering
\includegraphics[width=0.8\textwidth]{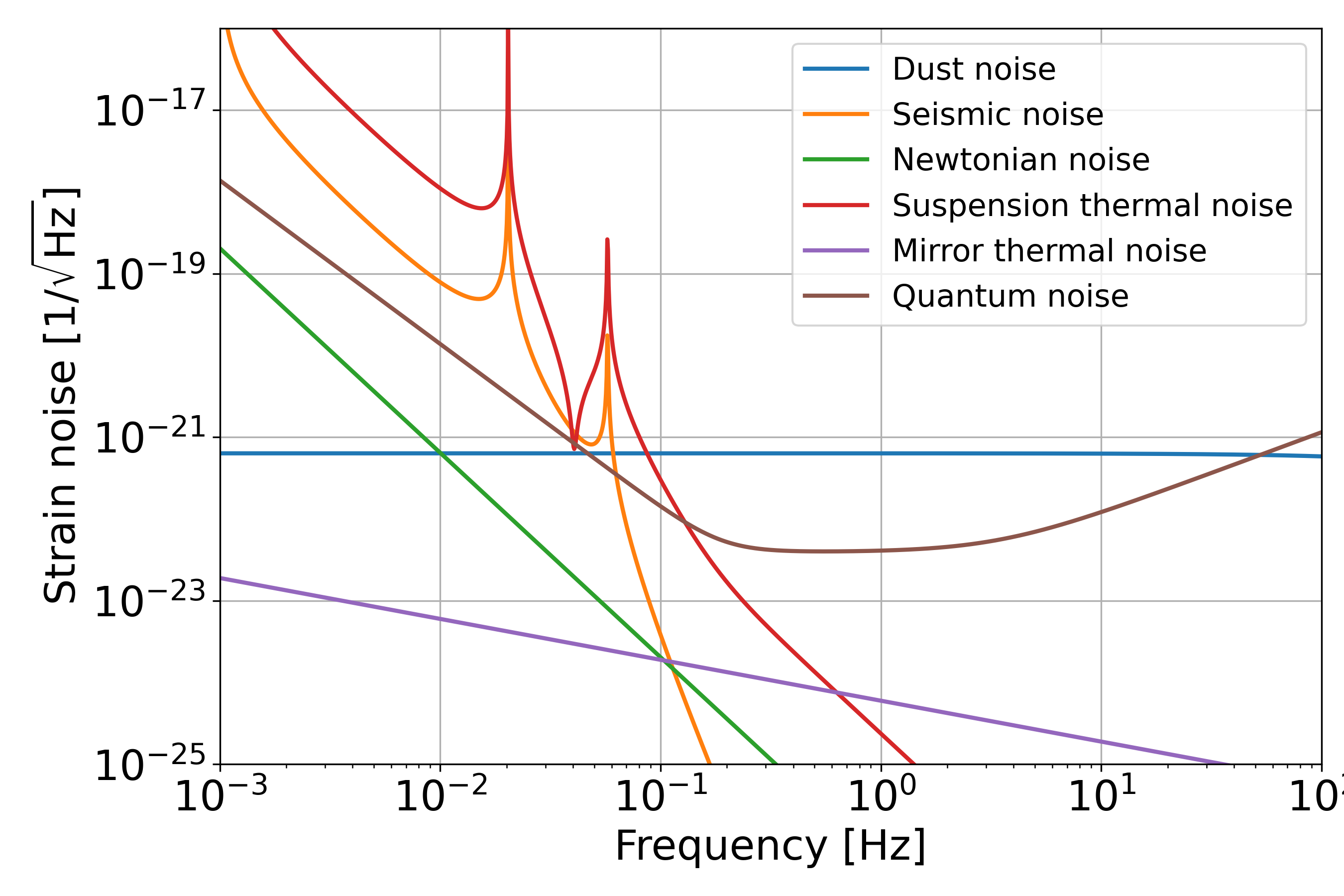}
\caption{Sensitivity models of a long-baseline laser-interferometric lunar GW detector. The suspension thermal noise requires a (today) unthinkable final suspension stage with resonance frequency at 1\,mHz (in horizontal and vertical direction) and quality factor $Q=10^9$ for the horizontal and $Q=10^5$ for the vertical.}
\label{fig:llisens}
\end{figure}

For what concerns the coating, the main characteristics to improve the noise budget are: coating deposition dimensions, materials and layer-layer and layer-substrate interfaces, that ultimately impact the value of the loss angle \cite{Penn_2003}. Another important parameter is the beam radius on the test masses, which can be regulated to some extent by the mirrors' radii of curvature. The larger the beam, the lower the mirror thermal noise. The price to pay is reduced stability of the arm cavity \cite{AcEA2020}. Suitable coating materials like amorphous silicon - fused silica for cryogenic interferometers are already being investigated for next-generation GW detectors \cite{AdEA2020,StMa2021}. They might provide a path to reach coating mechanical loss as low as $ \mathcal{O}(10^{-5})$ at cryogenic temperature. The choice of materials constraints the possible laser frequencies. For silicon substrates, a laser frequency of 1500\,nm can be chosen. The model parameter values are summarized in table \ref{tab:lbisustab}, and the corresponding noise budget is shown in figure \ref{fig:llisens}.

The noise budget does not contain noise from vertical filter stages. Since the vertical-to-horizontal coupling of a 35\,km long detector is about 1\%, it means that all noise from the suspensions along the vertical displacement direction cannot be more than a factor 100 larger compared to the displacement noise along the horizontal direction, which sets strong requirements on vertical seismic isolation and loss angles of the vertical filters. We derive three potential show-stoppers of the LBI-SUS concept from our noise models:
\begin{itemize}
    \item {\bf Suspension thermal noise}. We do not know today how to realize a horizontal suspension stage with resonance frequency of 1\,mHz and quality factor of $10^9$.
    \item {\bf Vertical seismic isolation}. We do not know today how to realize a vertical filter with 1\,mHz resonance frequency and quality factor of $10^5$. There is no damping dilution in vertical filters.
    \item {\bf Dust noise}. A pipe must be constructed to protect the beam from dust. Noise from stray light interacting with the pipe must be investigated.
\end{itemize}
Another technological challenge of the LBI-SUS concept, which does not appear in our modeled noise budget, is its angular controls especially with regards to the Sidles - Sigg effect. This effect results from optomechanical dynamics involving the suspended optics and laser beam \cite{SiSi2006,AnHa2021}. The optomechanical coupling gives rise to two eigenmodes for each of the two angular degrees of freedom of the two test masses forming an arm cavity. These modes are commonly known as soft and hard mode. The frequencies can be calculated using 
\beq
\begin{split}
\kappa &= \frac{2P_{\rm cav}L}{c(g_1g_2-1)},\\
\tau_{\rm soft} &= \frac{\kappa}{2}\left(g_1+g_2+\sqrt{(g_1-g_2)^2+4)}\right),\\
\tau_{\rm hard} &= \frac{\kappa}{2}\left(g_1+g_2-\sqrt{(g_1-g_2)^2+4)}\right),\\
f_{\rm soft} &= \frac{1}{2\pi}\sqrt{(2\pi f_{\rm mech})^2+\tau_{\rm soft}/\mathcal I},\\
f_{\rm hard} &= \frac{1}{2\pi}\sqrt{(2\pi f_{\rm mech})^2+\tau_{\rm hard}/\mathcal I},
\end{split}
\eeq
where $\mathcal I$ is the moment of inertia of the mirror around the rotation axis (e.g., a horizontal axis perpendicular to the direction of the arm in case of mirror pitch motion), and $P_{\rm cav}$ is the power of the laser beam inside the arm cavities. Assuming a mechanical angular mode of the suspended test mass at frequency $f_{\rm mech}=1$\,mHz, the coupled system has eigenfrequencies at $f_{\rm soft}=2.9\,$mHz and $f_{\rm hard}=4.8\,$mHz. Both mode frequencies lie well below the decihertz band helping the angular controls, but it is still important to model the noise introduced by the angular controls and all auxiliary controls, since these systems are known to introduce important low-frequency noise in terrestrial GW detectors \cite{MaEA2016}.

\section{Observational capabilities of lunar GW detectors}
\label{sec:capa}
The computation of noise budgets for the lunar detectors considered in this paper allows us to
estimate and compare their scientific capabilities. Several studies have already highlighted the importance of a decihertz observatory \cite{JaLo2021, SeEA2020, MSV2018, IsEA2018, Harms2021b}. Considering all concepts, lunar GW detection might cover the band from 1\,mHz to a few Hz and open the decihertz band to GW observations \cite{Har2022a}. It should be noted that in addition to the concepts proposed in recent years, additional considerations were made to enhance their science case. For example, antipodal pairs of GW detectors at the two lunar poles would form an ideal network for the search of stochastic GWs exploiting correlations between two detectors \cite{CoHa2014}. It was also pointed out that a distribution of vibration sensors over the surface of a sphere would enable detailed measurements of the GW polarization and interesting tests of general relativity \cite{WaPa1976,BiEA1996}. In this section, we will focus on the three main concepts as individual detectors, deployed in a single spot of the lunar surface, and leave the analysis of lunar GW detector networks to a future study.

\begin{figure}
\centering
\includegraphics[width=0.8\textwidth]{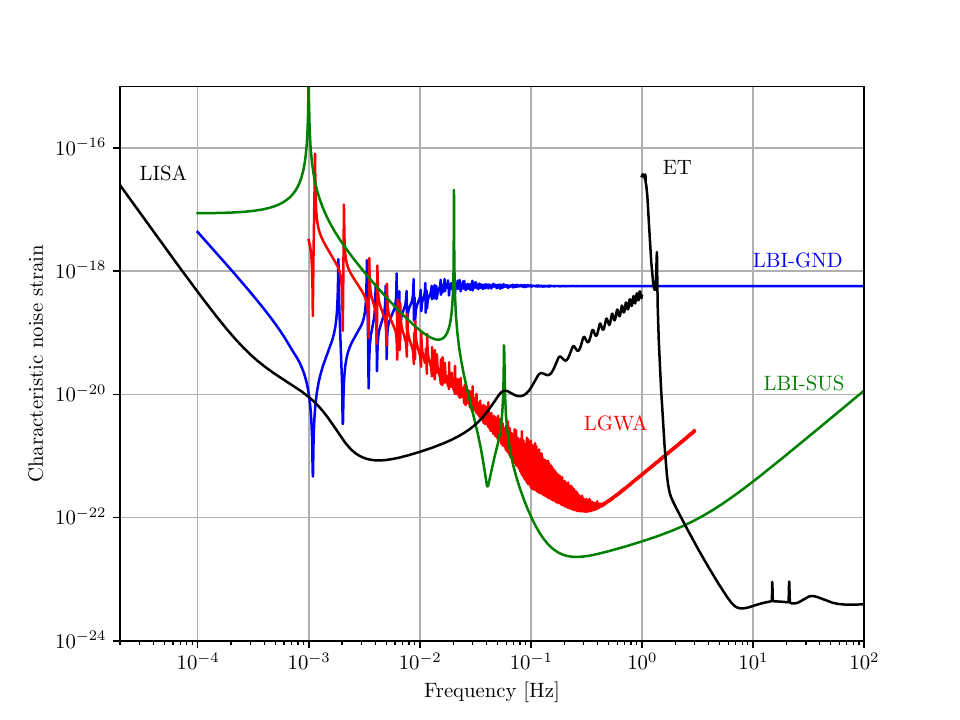}
\label{fig:sensitivities}
\caption{Estimated sensitivities of the detector concepts discussed in this work, compared with the ones of the Laser Interferometer Space Antenna \cite{Robson_2019} and the Einstein Telescope \cite{ET2020}. The LBI-SUS sensitivity is plotted without the dust noise assuming that a protective pipe is built around the laser beam.}
\end{figure}

\begin{figure}
\centering
\includegraphics[width=0.8\textwidth]{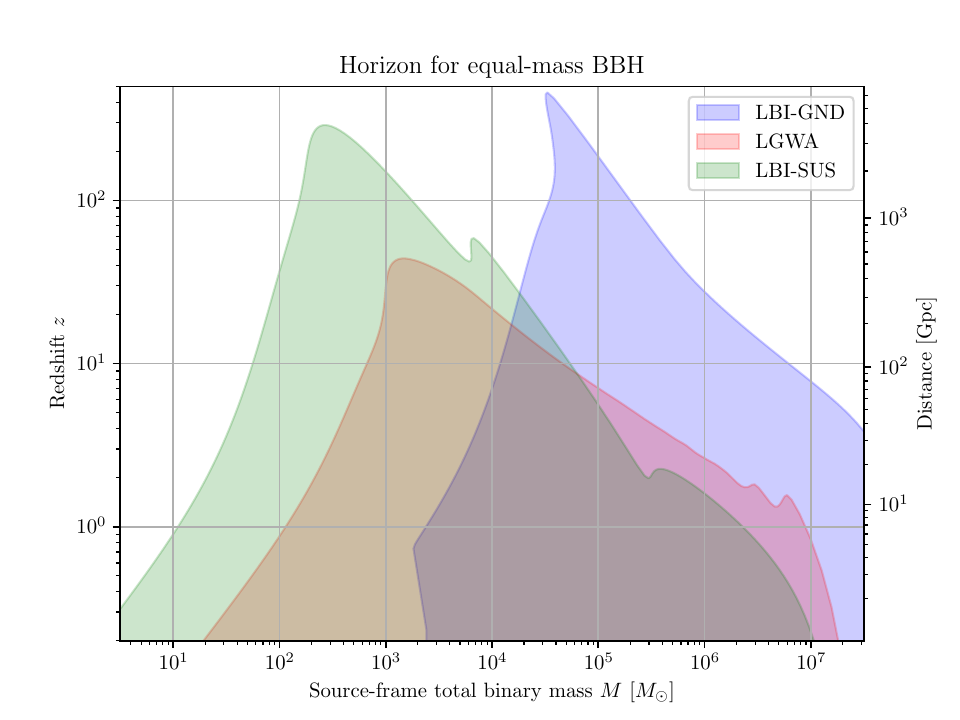}
\caption{Detection horizon for an optimally-oriented binary of black holes, at a signal-to-noise ratio of 10. The masses on the horizontal axis are in the source frame, but the waveform observed depends on the detector-frame mass, which is affected by redshift as $M_{\text{det}} = (1+z) M_{\text{source}}$. The distance shown on the vertical axis is luminosity distance, defined in a cosmological context as the quantity $d_L$ for which the flux scales as $F \propto 1/d_L^2$. The relation between luminosity distance and redshift is computed assuming a cosmology consistent with the Planck mission's 2018 results \cite{Planck2018}.}
\label{fig:horizon}
\end{figure}

For the LBI-SUS analysis, we assume that the dust noise (section \ref{sec:dust}) is removed by a shielding pipe. Without this shield, we think that the LBI-SUS sensitivity would not be good enough to justify the deployment of such a complex instrument on the Moon. Figure \ref{fig:sensitivities} shows the noise curves of all three concepts expressed in terms of characteristic strain noise, $h_n (f) = \sqrt{f S_n(f)}$. This dimensionless quantity is useful since it can be compared on a logarithmic axis with the characteristic signal strain, $h_s(f) = 2 f |\widetilde{h}(f)|$, to yield the signal to noise ratio (SNR):
\beq 
\text{SNR}^2 = \int \frac{h_s^2 (f)}{h_n^2 (f)} \text{d}\log f \,.
\eeq
LBI-SUS and LGWA are candidates for the decihertz band, while the LBI-GND concept is most sensitive in the millihertz band, and it would not provide an improvement in the decihertz band if LISA reaches its design sensitivity.

The emission frequency of compact object binaries is inversely proportional to their mass; the ones 
that have been detected so far by ground-based interferometers are all in the range of one to several tens of solar masses. The binaries merging in the decihertz band, therefore, would be on the order of thousands of solar masses: the very center of  the Intermediate Mass Black Hole (IMBH) range, $10^2$ to $10^5 M_\odot$ \cite{SeEA2020}. The amplitude of the emission also scales linearly with the mass, meaning that even if very distant, these binaries could still be detectable. 

\begin{figure}
\centering
\includegraphics[width=0.8\textwidth]{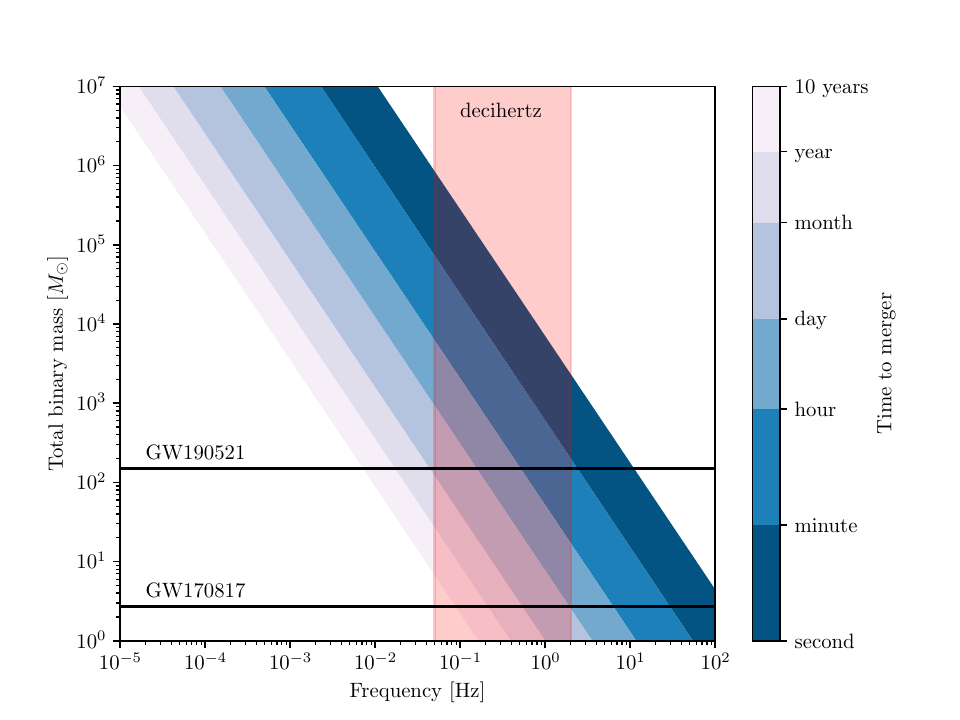}
\caption{Time to merger for a compact binary as a function of total mass (in the detector frame) and frequency. We highlight the decihertz band from 0.05Hz to 2Hz, where lunar detectors such as LGWA and LBI-SUS have an edge over space- and ground-based ones, and report for reference the total mass of the most and least massive events to date, GW190521 and GW170817 (for the latter, that is the total mass the binary would have had if the mass ratio was 1). These times are computed with a lowest-order post-Newtonian approximation, which is very accurate at low frequencies but imprecise at the high frequency end.}
\label{fig:time-to-merger}
\end{figure}

\begin{figure}
\centering
\includegraphics[width=0.8\textwidth]{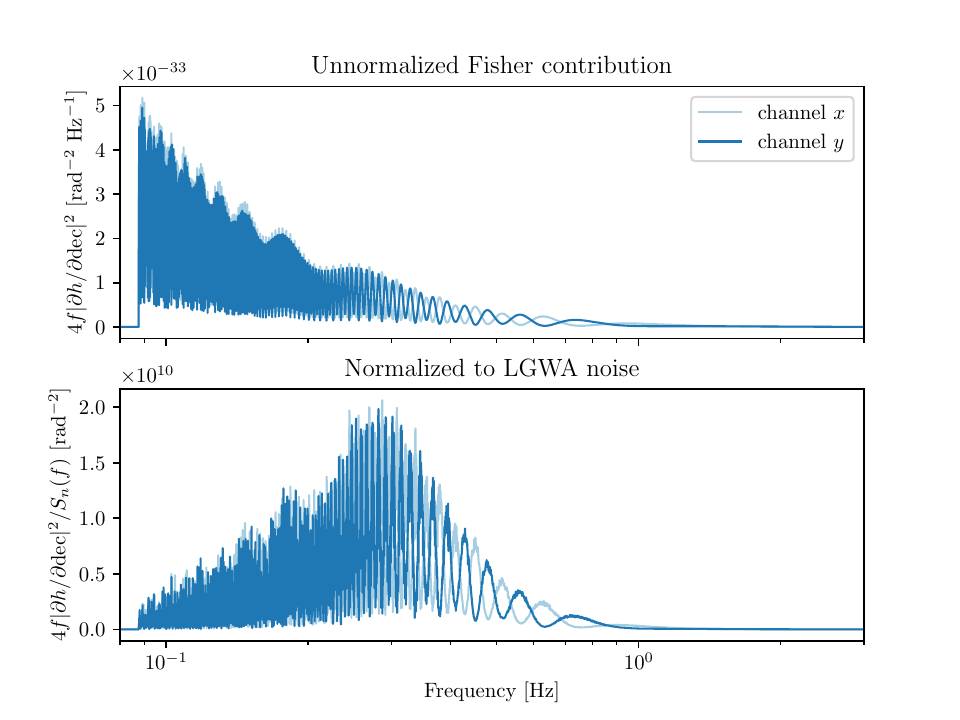}
\caption{Fisher matrix integrand for the dependence of the signal on declination. The signal is a neutron star binary with parameters set to their median values for the GW170817 detection \cite{AbEA2017d}. In the upper panel, we show the square modulus of the derivative of the waveform with respect to declination, multiplied by the frequency $f$. This does not depend on the noise PSD, but only on the antenna pattern and the motion of the Moon. In the bottom panel, we show this quantity normalized to the estimated sensitivity of LGWA. For this analysis, we fix the orbital phase term at the upper edge of the detector band, $\exp(-\mathrm{i} \vec{k} \cdot \vec{r})$, to zero --- this is an arbitrary phase choice and has no impact on the final localization, but without it the integrand would have a large contribution at high frequency, which is however irrelevant due to being completely degenerate with the arrival time parameter.
With a 10-year observation, this signal would be localized to within 3 arcmin$^2$ (90\% area). The oscillations in the Fisher contribution are monthly and yearly: the largest contributor to the localization is the last year.}
\label{fig:fisher-integrand}
\end{figure}

In order to compute the distances to which these could be detected, shown in figure \ref{fig:horizon}, we need to make an assumption on how the gravitational strain tensor $h_{ij}$ maps onto the strain measured by the sensor, $h(t)$.
We consider a linear response to these waves given by $h = h_{ij} D_{ij}$, where the detection tensor is
\beq
D_{ij}^{\text{LBI}} = \frac{1}{2} \left( a_i a_j - b_i b_j\right)
\eeq
for interferometric detectors, where $a$ and $b$ are two unit vectors along the interferometer's arms (which we assume to be orthogonal and located near a lunar pole), while
\beq
D_{ij}^{\text{LGWA}} = n_i b_j
\eeq
for seismic inertial measurement \cite{DuEA2022}, where $n$ is the normal unit vector to the surface at the location of the sensor, while $b_j$ is a tangent unit vector in the direction where the surface displacement is being measured. We assume to have 4 stations (resulting in an equivalent reduction in the PSD by a factor 4) and two measurement channels per station in orthogonal directions.

For the signals we are considering (without effects such as eccentricity or precession), we can compute a one-to-one function relating frequency and time of arrival, shown schematically in figure \ref{fig:time-to-merger}. This allows us to compute the varying antenna pattern as a function of frequency. We assume a lifetime of 10 years for all detectors, and truncate the signals which would last longer than this in the low-frequency end --- this is the optimal observation scenario, in which the signal comes in band as soon as the detector is turned on, and ends right at the end of the detector lifetime.

There is a phase effect due to the motion of the Moon around the Sun, which in the frequency domain can be expressed as
\beq
\phi = - \vec{k} \cdot \vec{r}_\text{Moon}\,,
\eeq
where $\vec{k} = (2 \pi f / c) \hat{k}$ is the wave vector of the incoming gravitational wave (which we consider as fixed in our solar-centered International Celestial Reference System (ICRS) \cite{SOFA:2021-01-25}), while $\vec{r}_\text{Moon}$ is the position of the detector, which we compute using the astropy \cite{astropy:2022} interface to jplephem \cite{pyephem2011}.

Figure \ref{fig:horizon} shows the distances to which equal-mass black hole binaries could be detected if they were optimally oriented. For all detectors, these are cosmological distances, reaching the formation of the earliest galaxies. Little is known about IMBHs, and a decihertz observatory would either detect their mergers, or put an extremely strong constraint on their population. Solar-mass binaries are also a target of interest for LGWA and LBI-SUS, even though the distances to which they could be detected are not as large. These would be detected months to years before their merger, which means that during their orbit the Moon would cover a significant distance in its orbit around the Earth and in turn around the Sun. This would correspond to a large phase term in the observed waveform: the relevant quantities to compare to estimate it are the wavelength of the gravitational wave, ranging from 1 to 10 light-seconds in the decihertz band, and the Moon-Sun orbital diameter, which is around 1000 light-seconds. The phase due to the orbit of the Moon around the Earth is in general less significant, on the order of a few cycles at most at the upper range of the detector band, but it still gives a contribution due to its faster evolution.

While a comprehensive analysis on the localization capabilities of such detectors has not been yet performed, we can give some estimate using the Fisher matrix formalism. Specifically, we can compute with GWFish \cite{DuEA2022} the Fisher integrand for the dependence of the Fourier-domain signal $h(f)$ on one of the two angles defining the signal position, such as declination: 
\beq
F_{\text{dec, dec}} = \int \frac{4}{S_n(f)} \left|\frac{\partial h}{\partial \text{dec}}\right|^2 \text{d}f
= \int \underbrace{\frac{4f}{S_n(f)} \left|\frac{\partial h}{\partial \text{dec}}\right|^2}_{\text{Fisher integrand}} \text{d} \log f \,.
\eeq

This quantifies the extent to which each frequency region contributes to the estimation of the parameter, in the Gaussian approximation. In figure \ref{fig:fisher-integrand}, we show the integrand in $\text{d}\log f$ as opposed to the one in $\text{d}f$, so that the Fisher matrix element is visually represented by the area of the curve being shown on a logarithmic axis.

We consider a signal with the same parameters as GW170817 as seen by the LGWA detector, where it would have an SNR of 32, very similar to the one it had for the LVK network. We can see modulations with both the period of the Earth-Moon system's orbit around the Sun, and the period of the rotation of the Moon. The full Fisher analysis requires us to compute all the Fisher matrix elements and invert the matrix, thus accounting for correlations; it yields a 90\% sky area of roughly 3 arcmin$^2$.

The upper panel of figure \ref{fig:fisher-integrand} shows the extent to which rotation around the Sun can contribute to the localization of a binary that stays in band for a long time. It does not depend on the specifics of the lunar detector as long as it is sensitive enough to observe the source. Repeating the same analysis for the LBI-SUS detector yields analogous results, and an even smaller sky localization area due to the higher SNR.

The decihertz band is crucial to this end, since the orbital contribution to the localization capabilities is the largest there; it would be even larger at millihertz frequencies, but stellar-mass binaries detected in that band will not merge for hundreds or thousands of years, meaning that while we could localize them, this would likely not be connected with any multimessenger observation since we could not follow them all the way to merger. Therefore, the decihertz band has a unique potential for precise localization, which can only be matched by several ground-based detectors with high SNR.

\section{Conclusion}
We presented noise models of lunar GW detector concepts. This includes LGWA as well as long-baseline, laser-interferometric concepts. Laser interferometers can oberve GWs either with optics mounted to the ground (LBI-GND) or with suspended test masses (LBI-SUS). We find that the LBI-GND concept can achieve excellent sensitivity in the millihertz band without technological show-stoppers; at least based on our study of the main instrumental noises. The LBI-SUS concept has the potential to become a ground-breaking decihertz GW observatory, but there are a few potential show-stoppers that must be addressed, which include dust noise, suspension thermal noise, and seismic isolation for vertical surface displacement. 

It is unclear whether an ambitious LBI-SUS concept can ever be realized on the Moon, or whether it has any crucial advantages over space-based decihertz concepts. However, the extremely low seismic background on the Moon is an enormous advantage compared to terrestrial detectors. One of the main show-stoppers for decihertz GW observations with terrestrial detectors, the gravitational background noise produced for example by the seismic field, would play a minor role for a lunar LBI-SUS detector. The LGWA is a much less complex system and relies mostly on technologies that are within reach. A potential role of the LBI-GND concept could be to realize a long-lived observatory in the millihertz band; albeit with much lower sensitivity than LISA. Both, LBI-GND and LGWA, would also be outstanding lunar geophysical observatories. 

All three technological approaches (LGWA, LBI-GND, LBI-SUS) have the potential to detect GW sources out to redshifts of $z=30$ and beyond. Many publications have already emphasized the enormous potential for breakthrough science with GW observations in the decihertz band. The Moon certainly has the potential to become the third impactful platform for GW detectors together with Earth and space.

\begin{acknowledgments}
This work is the outcome of a lecture series at GSSI on lunar GW detection, and we would like to express our gratitude for the support of the GSSI administration and of the coordinator of the PhD lecture program.
Also, JT would like to thank Tito Dal Canton for a fruitful discussion on the localization capabilities of a lunar detector.
\end{acknowledgments}



\bibliographystyle{newRS}

\bibliography{references.bib}

\begin{thebibliography}{99}

\bibitem{GiEA1977}
Giganti, J., Larson, J. et~al. 1977  {Lunar surface gravimeter experiment. Final report}. Technical report University of Maryland Department of Physics and Astronomy, College Park, Md.

\bibitem{NaEA1981}
Nakamura, Y., Latham, G.V. et~al. 1981  {Passive Seismic Experiment, Long Period Event Catalog, Final Version (1969 Day 202-1977 Day 273, ALSEP Stations 11, 12, 13, 14, 15, and 16)}. Technical report University of Texas Institute for Geophysics.

\bibitem{LoEA2009}
Lognonné, P., Le~Feuvre, M. et~al. 2009  {Moon meteoritic seismic hum: Steady state prediction}. {\em Journal of Geophysical Research: Planets} \textbf{114}.
(\href{http://dx.doi.org/10.1029/2008JE003294}{10.1029/2008JE003294})

\bibitem{WiEA2019}
Williams, J.P., Greenhagen, B.T. et~al. 2019  Seasonal Polar Temperatures on the Moon. {\em Journal of Geophysical Research: Planets} \textbf{124}, 2505--2521.
(\href{http://dx.doi.org/10.1029/2019JE006028}{10.1029/2019JE006028})

\bibitem{HaEA2021a}
Harms, J., Ambrosino, F. et~al. 2021  Lunar Gravitational-wave Antenna. {\em The Astrophysical Journal} \textbf{910}, 1.
(\href{http://dx.doi.org/10.3847/1538-4357/abe5a7}{10.3847/1538-4357/abe5a7})

\bibitem{KaEA2020}
Katsanevas, S., Bernard, P. et~al. 2020  {Lunar Seismic and Gravitational Antenna}. In {\em {Ideas for exploring the Moon with a large European lander}} ,  p.~1. ESA.

\bibitem{JaLo2021}
Jani, K. and Loeb, A.. 2021  {Gravitational-wave Lunar Observatory for Cosmology}. {\em Journal of Cosmology and Astroparticle Physics} \textbf{2021}, 044.
(\href{http://dx.doi.org/10.1088/1475-7516/2021/06/044}{10.1088/1475-7516/2021/06/044})

\bibitem{ASEA2021}
Amaro-Seoane, P., Bischof, L. et~al. 2021  {LION: laser interferometer on the moon}. {\em Classical and Quantum Gravity} \textbf{38}, 125008.
(\href{http://dx.doi.org/10.1088/1361-6382/abf441}{10.1088/1361-6382/abf441})

\bibitem{LiEA2023}
Li, J., Liu, F. et~al. 2023  Detecting gravitational wave with an interferometric seismometer array on lunar nearside. {\em SCIENCE CHINA Physics, Mechanics \& Astronomy} \textbf{66}, 109513--.
(\href{http://dx.doi.org/10.1007/s11433-023-2179-9}{10.1007/s11433-023-2179-9})

\bibitem{AcEA2015}
Acernese, F., Agathos, M. et~al. 2014  Advanced Virgo: a second-generation interferometric gravitational wave detector. {\em Classical and Quantum Gravity} \textbf{32}, 024001.
(\href{http://dx.doi.org/10.1088/0264-9381/32/2/024001}{10.1088/0264-9381/32/2/024001})

\bibitem{LSC2015}
Aasi, J., Abbott, B.P. et~al. 2015  {Advanced LIGO}. {\em Classical and Quantum Gravity} \textbf{32}, 074001.
(\href{http://dx.doi.org/10.1088/0264-9381/32/7/074001}{10.1088/0264-9381/32/7/074001})

\bibitem{Har2022a}
Harms, J.. 2022  Seismic Background Limitation of Lunar Gravitational-Wave Detectors. {\em Phys. Rev. Lett.} \textbf{129}, 071102.
(\href{http://dx.doi.org/10.1103/PhysRevLett.129.071102}{10.1103/PhysRevLett.129.071102})

\bibitem{ASEA2017}
{Amaro-Seoane}, P., {Audley}, H. et~al. 2017  {Laser Interferometer Space Antenna}. {\em arXiv e-prints}.

\bibitem{ET2020}
{ET Steering Committee}. 2020  {Einstein Telescope design report update 2020}. {\em {available from European Gravitational Observatory, document number ET-0007B-20}}.

\bibitem{CuHo2009}
Cutler, C. and Holz, D.E.. 2009  Ultrahigh precision cosmology from gravitational waves. {\em Phys. Rev. D} \textbf{80}, 104009.
(\href{http://dx.doi.org/10.1103/PhysRevD.80.104009}{10.1103/PhysRevD.80.104009})

\bibitem{MSV2018}
Mandel, I., Sesana, A. and Vecchio, A.. 2018  The astrophysical science case for a decihertz gravitational-wave detector. {\em Classical and Quantum Gravity} \textbf{35}, 054004.
(\href{http://dx.doi.org/10.1088/1361-6382/aaa7e0}{10.1088/1361-6382/aaa7e0})

\bibitem{JSC2019}
Jani, K., Shoemaker, D. and Cutler, C.. 2020  Detectability of intermediate-mass black holes in multiband gravitational wave astronomy. {\em Nature Astronomy} \textbf{4}, 260--265.
(\href{http://dx.doi.org/10.1038/s41550-019-0932-7}{10.1038/s41550-019-0932-7})

\bibitem{SeEA2020}
Arca~Sedda, M., Berry, C.P.L. et~al. 2020  The missing link in gravitational-wave astronomy: discoveries waiting in the decihertz range. {\em Classical and Quantum Gravity} \textbf{37}, 215011.
(\href{http://dx.doi.org/10.1088/1361-6382/abb5c1}{10.1088/1361-6382/abb5c1})

\bibitem{Phi2003}
{Phinney et al.}. 2003  The Big Bang Observer: direct detection of gravitational waves from the birth of the universe to the present. {\em {NASA Mission Concept Study}}.

\bibitem{KaEA2021}
Kawamura, S., Ando, M. et~al. 2021  {Current status of space gravitational wave antenna DECIGO and B-DECIGO}. {\em Progress of Theoretical and Experimental Physics} \textbf{2021}.
(\href{http://dx.doi.org/10.1093/ptep/ptab019}{10.1093/ptep/ptab019})

\bibitem{CuHa2006}
Cutler, C. and Harms, J.. 2006  {Big Bang Observer and the neutron-star-binary subtraction problem}. {\em Phys. Rev. D} \textbf{73}, 042001.
(\href{http://dx.doi.org/10.1103/PhysRevD.73.042001}{10.1103/PhysRevD.73.042001})

\bibitem{HaEA2008}
Harms, J., Mahrdt, C. et~al. 2008  Subtraction-noise projection in gravitational-wave detector networks. {\em Phys. Rev. D} \textbf{77}, 123010.
(\href{http://dx.doi.org/10.1103/PhysRevD.77.123010}{10.1103/PhysRevD.77.123010})

\bibitem{vHEA2023}
van Heijningen, J., ter Brake, M. et~al. 2023  {The payload of the Lunar Gravitational-wave Antenna}. .

\bibitem{EvEA2021}
Evans, M., Adhikari, R.X. et~al. 2021  A Horizon Study for Cosmic Explorer: Science, Observatories, and Community. .

\bibitem{DuSu1974}
Duennebier, F. and Sutton, G.H.. 1974  Thermal moonquakes. {\em Journal of Geophysical Research (1896-1977)} \textbf{79}, 4351--4363.
(\href{http://dx.doi.org/https://doi.org/10.1029/JB079i029p04351}{https://doi.org/10.1029/JB079i029p04351})

\bibitem{BERTOLINI2006616}
Bertolini, A., DeSalvo, R. et~al. 2006a  Mechanical design of a single-axis monolithic accelerometer for advanced seismic attenuation systems. {\em Nuclear Instruments and Methods in Physics Research Section A: Accelerators, Spectrometers, Detectors and Associated Equipment} \textbf{556}, 616--623.
(\href{http://dx.doi.org/https://doi.org/10.1016/j.nima.2005.10.117}{https://doi.org/10.1016/j.nima.2005.10.117})

\bibitem{BeEA2006}
Bertolini, A., DeSalvo, R. et~al. 2006b  Mechanical design of a single-axis monolithic accelerometer for advanced seismic attenuation systems. {\em Nuclear Instruments and Methods in Physics Research Section A: Accelerators, Spectrometers, Detectors and Associated Equipment} \textbf{556}, 616 -- 623.
(\href{http://dx.doi.org/https://doi.org/10.1016/j.nima.2005.10.117}{https://doi.org/10.1016/j.nima.2005.10.117})

\bibitem{Ferreira_2021}
Ferreira, E., Bocchese, F. et~al. 2021  Superconducting thin film spiral coils as low-noise cryogenic actuators. {\em Journal of Physics: Conference Series} \textbf{2156}, 012080.
(\href{http://dx.doi.org/10.1088/1742-6596/2156/1/012080}{10.1088/1742-6596/2156/1/012080})

\bibitem{BiEA2002}
Bilenko, I.A., Ju, L. et~al. 2002  {Niobium flexure suspension design for high Q sapphire test masses for future gravitational wave detectors}. {\em Measurement Science and Technology} \textbf{13}, 1173.
(\href{http://dx.doi.org/10.1088/0957-0233/13/8/302}{10.1088/0957-0233/13/8/302})

\bibitem{EcGe2022}
Eckhardt, T. and Gerberding, O.. 2022  {Noise Limitations in Multi-Fringe Readout of Laser Interferometers and Resonators}. {\em Metrology} \textbf{2}, 98--113.
(\href{http://dx.doi.org/10.3390/metrology2010007}{10.3390/metrology2010007})

\bibitem{Ben1983}
Ben-Menahem, A.. 1983  {Excitation of the Earth's eigenvibrations by gravitational radiation from astrophysical sources}. {\em Il Nuovo Cimento C} \textbf{6}, 49--71.
(\href{http://dx.doi.org/10.1007/BF02511372}{10.1007/BF02511372})

\bibitem{CoHa2014b}
Coughlin, M. and Harms, J.. 2014  {Constraining the gravitational wave energy density of the Universe using Earth's ring}. {\em Phys. Rev. D} \textbf{90}, 042005.
(\href{http://dx.doi.org/10.1103/PhysRevD.90.042005}{10.1103/PhysRevD.90.042005})

\bibitem{GaEA2019}
Garcia, R.F., Khan, A. et~al. 2019  Lunar Seismology: An Update on Interior Structure Models. {\em Space Science Reviews} \textbf{215}, 50.
(\href{http://dx.doi.org/10.1007/s11214-019-0613-y}{10.1007/s11214-019-0613-y})

\bibitem{LaEA1970}
Latham, G.V., Ewing, M. et~al. 1970  Passive Seismic Experiment. {\em Science} \textbf{167}, 455--457.
(\href{http://dx.doi.org/10.1126/science.167.3918.455}{10.1126/science.167.3918.455})

\bibitem{BuCh2001}
Buonanno, A. and Chen, Y.. 2001  Quantum noise in second generation, signal-recycled laser interferometric gravitational-wave detectors. {\em Phys. Rev. D} \textbf{64}, 042006.
(\href{http://dx.doi.org/10.1103/PhysRevD.64.042006}{10.1103/PhysRevD.64.042006})

\bibitem{BHS2018}
Barsotti, L., Harms, J. and Schnabel, R.. 2018  Squeezed vacuum states of light for gravitational wave detectors. {\em Reports on Progress in Physics} \textbf{82}, 016905.
(\href{http://dx.doi.org/10.1088/1361-6633/aab906}{10.1088/1361-6633/aab906})

\bibitem{DaEA2018}
Daubar, I., Lognonn{\'e}, P. et~al. 2018  Impact-Seismic Investigations of the InSight Mission. {\em Space Science Reviews} \textbf{214}, 132.
(\href{http://dx.doi.org/10.1007/s11214-018-0562-x}{10.1007/s11214-018-0562-x})

\bibitem{PaEA2022}
{Panning}, M.P., {Kedar}, S. et~al. 2022  {Farside Seismic Suite (FSS): Surviving the Lunar Night and Delivering the First Seismic Data from the Farside of the Moon}. In {\em 53rd Lunar and Planetary Science Conference} vol. 2678{\em LPI Contributions} p. 1576.

\bibitem{MaEA2015}
Matichard, F., Lantz, B. et~al. 2015  {Advanced LIGO two-stage twelve-axis vibration isolation and positioning platform. Part 2: Experimental investigation and tests results}. {\em Precision Engineering} \textbf{40}, 287 -- 297.
(\href{http://dx.doi.org/http://dx.doi.org/10.1016/j.precisioneng.2014.11.010}{http://dx.doi.org/10.1016/j.precisioneng.2014.11.010})

\bibitem{AcEA2010}
Acernese, F., Antonucci, F. et~al. 2010  {Measurements of Superattenuator seismic isolation by Virgo interferometer}. {\em Astroparticle Physics} \textbf{33}, 182 -- 189.
(\href{http://dx.doi.org/http://dx.doi.org/10.1016/j.astropartphys.2010.01.006}{http://dx.doi.org/10.1016/j.astropartphys.2010.01.006})

\bibitem{Win2002}
Winterflood, J.. 2002 {\em High performance vibration isolation for gravitational wave detection}.
PhD thesis University of Western Australia.

\bibitem{SaEA1994}
Saulson, P.R., Stebbins, R.T. et~al. 1994  The inverted pendulum as a probe of anelasticity. {\em Review of Scientific Instruments} \textbf{65}, 182--191.
(\href{http://dx.doi.org/10.1063/1.1144774}{10.1063/1.1144774})

\bibitem{WLB1999}
Winterflood, J., Losurdo, G. and Blair, D.. 1999  Initial results from a long-period conical pendulum vibration isolator with application for gravitational wave detection. {\em Physics Letters A} \textbf{263}, 9 -- 14.
(\href{http://dx.doi.org/http://dx.doi.org/10.1016/S0375-9601(99)00715-X}{http://dx.doi.org/10.1016/S0375-9601(99)00715-X})

\bibitem{vHEA2023a}
{van Heijningen}, J., Winterflood, J. et~al. 2023  Multi-blade monolithic Euler springs with optimised stress distribution. {\em Journal of Sound and Vibration} \textbf{552}, 117614.
(\href{http://dx.doi.org/https://doi.org/10.1016/j.jsv.2023.117614}{https://doi.org/10.1016/j.jsv.2023.117614})

\bibitem{Har2019}
Harms, J.. 2019  Terrestrial gravity fluctuations. {\em Living Reviews in Relativity} \textbf{22}, 6.
(\href{http://dx.doi.org/10.1007/s41114-019-0022-2}{10.1007/s41114-019-0022-2})

\bibitem{Cel2000}
Cella, G.. 2000  {Off-Line Subtraction of Seismic Newtonian Noise}. In Casciaro, B., Fortunato, D. et~al, editors, {\em Recent Developments in General Relativity} ,  pp. 495--503. Springer Milan.
(\href{http://dx.doi.org/10.1007/978-88-470-2113-6\_44}{10.1007/978-88-470-2113-6\_44})

\bibitem{BaEA2020}
Badaracco, F., Harms, J. et~al. 2020  {Machine learning for gravitational-wave detection: surrogate Wiener filtering for the prediction and optimized cancellation of Newtonian noise at Virgo}. {\em Classical and Quantum Gravity} \textbf{37}, 195016.
(\href{http://dx.doi.org/10.1088/1361-6382/abab64}{10.1088/1361-6382/abab64})

\bibitem{HaML2018}
Harms, J. and Mow-Lowry, C.M.. 2017  Suspension-thermal noise in spring-antispring systems for future gravitational-wave detectors. {\em Classical and Quantum Gravity} \textbf{35}, 025008.
(\href{http://dx.doi.org/10.1088/1361-6382/aa9e28}{10.1088/1361-6382/aa9e28})

\bibitem{CaEA2000}
Cagnoli, G., Hough, J. et~al. 2000  Damping dilution factor for a pendulum in an interferometric gravitational waves detector. {\em Physics Letters A} \textbf{272}, 39 -- 45.
(\href{http://dx.doi.org/http://dx.doi.org/10.1016/S0375-9601(00)00411-4}{http://dx.doi.org/10.1016/S0375-9601(00)00411-4})

\bibitem{HaEA2002}
Harry, G.M., Gretarsson, A.M. et~al. 2002  Thermal noise in interferometric gravitational wave detectors due to dielectric optical coatings. {\em Classical and Quantum Gravity} \textbf{19}, 897.
(\href{http://dx.doi.org/10.1088/0264-9381/19/5/305}{10.1088/0264-9381/19/5/305})

\bibitem{EvEA2008}
Evans, M., Ballmer, S. et~al. 2008  Thermo-optic noise in coated mirrors for high-precision optical measurements. {\em Phys. Rev. D} \textbf{78}, 102003.
(\href{http://dx.doi.org/10.1103/PhysRevD.78.102003}{10.1103/PhysRevD.78.102003})

\bibitem{Lev1998a}
Levin, Y.. 1998  {Internal thermal noise in the LIGO test masses: A direct approach}. {\em Phys. Rev. D} \textbf{57}, 659--663.
(\href{http://dx.doi.org/10.1103/PhysRevD.57.659}{10.1103/PhysRevD.57.659})

\bibitem{HoEA2013}
Hong, T., Yang, H. et~al. 2013  Brownian thermal noise in multilayer coated mirrors. {\em Phys. Rev. D} \textbf{87}, 082001.
(\href{http://dx.doi.org/10.1103/PhysRevD.87.082001}{10.1103/PhysRevD.87.082001})

\bibitem{NaEA2002}
Nakagawa, N., Gretarsson, A.M. et~al. 2002  Thermal noise in half-infinite mirrors with nonuniform loss: A slab of excess loss in a half-infinite mirror. {\em Phys. Rev. D} \textbf{65}, 102001.
(\href{http://dx.doi.org/10.1103/PhysRevD.65.102001}{10.1103/PhysRevD.65.102001})

\bibitem{AdEA2020}
Adhikari, R.X., Arai, K. et~al. 2020  A cryogenic silicon interferometer for gravitational-wave detection. {\em Classical and Quantum Gravity} \textbf{37}, 165003.
(\href{http://dx.doi.org/10.1088/1361-6382/ab9143}{10.1088/1361-6382/ab9143})

\bibitem{RPH2009}
Rubanu, F., Poggiani, R. and Hough, J.. 2009  Interplanetary dust: a source of noise for LISA?. {\em Classical and Quantum Gravity} \textbf{26}, 225012.
(\href{http://dx.doi.org/10.1088/0264-9381/26/22/225012}{10.1088/0264-9381/26/22/225012})

\bibitem{HoEA2015}
Hor{\'a}nyi, M., Szalay, J.R. et~al. 2015  {A permanent, asymmetric dust cloud around the Moon}. {\em Nature} \textbf{522}, 324--326.
(\href{http://dx.doi.org/10.1038/nature14479}{10.1038/nature14479})

\bibitem{SzEA2019}
Szalay, J.R., Pokorný, P. et~al. 2019  Impact Ejecta and Gardening in the Lunar Polar Regions. {\em Journal of Geophysical Research: Planets} \textbf{124}, 143--154.
(\href{http://dx.doi.org/https://doi.org/10.1029/2018JE005756}{https://doi.org/10.1029/2018JE005756})

\bibitem{Hul1981}
van~de Hulst, H.C.. 1981 {\em Light scattering by small particles}.
Dover Publication, Inc.

\bibitem{NaEA2008}
Nawrodt, R., Zimmer, A. et~al. 2008  {High mechanical Q-factor measurements on silicon bulk samples}. {\em Journal of Physics: Conference Series} \textbf{122}, 012008.
(\href{http://dx.doi.org/10.1088/1742-6596/122/1/012008}{10.1088/1742-6596/122/1/012008})

\bibitem{Sch2008}
Schr{\"o}ter, A.. 2008 {\em Mechanical losses in materials for future cryogenic gravitational wave detectors}.
PhD thesis Friedrich-Schiller-Universität Jena.

\bibitem{Tin2014}
Tinto, M. and Dhurandhar, S.V.. 2014  Time-Delay Interferometry. {\em Living Reviews in Relativity} \textbf{17}, 6.
(\href{http://dx.doi.org/10.12942/lrr-2014-6}{10.12942/lrr-2014-6})

\bibitem{HiEA2022}
Hilweg, C., Shadmany, D. et~al. 2022  Limits and prospects for long-baseline optical fiber interferometry. {\em Optica} \textbf{9}, 1238--1252.
(\href{http://dx.doi.org/10.1364/OPTICA.470430}{10.1364/OPTICA.470430})

\bibitem{varvella2004}
Varvella, M., Calloni, E. et~al. 2004  Feasibility of a magnetic suspension for second generation gravitational wave interferometers. {\em Astroparticle Physics} \textbf{21}, 325--335.
(\href{http://dx.doi.org/https://doi.org/10.1016/j.astropartphys.2004.01.002}{https://doi.org/10.1016/j.astropartphys.2004.01.002})

\bibitem{Penn_2003}
Penn, S.D., Sneddon, P.H. et~al. 2003  Mechanical loss in tantala/silica dielectric mirror coatings. {\em Classical and Quantum Gravity} \textbf{20}, 2917.
(\href{http://dx.doi.org/10.1088/0264-9381/20/13/334}{10.1088/0264-9381/20/13/334})

\bibitem{AcEA2020}
Acernese, F., Agathos, M. et~al. 2020  {The advanced Virgo longitudinal control system for the O2 observing run}. {\em Astroparticle Physics} \textbf{116}, 102386.
(\href{http://dx.doi.org/https://doi.org/10.1016/j.astropartphys.2019.07.005}{https://doi.org/10.1016/j.astropartphys.2019.07.005})

\bibitem{StMa2021}
Steinlechner, J. and Martin, I.W.. 2021  How can amorphous silicon improve current gravitational-wave detectors?. {\em Phys. Rev. D} \textbf{103}, 042001.
(\href{http://dx.doi.org/10.1103/PhysRevD.103.042001}{10.1103/PhysRevD.103.042001})

\bibitem{SiSi2006}
Sidles, J.A. and Sigg, D.. 2006  {Optical torques in suspended Fabry-Perot interferometers}. {\em Physics Letters A} \textbf{354}, 167 -- 172.
(\href{http://dx.doi.org/http://dx.doi.org/10.1016/j.physleta.2006.01.051}{http://dx.doi.org/10.1016/j.physleta.2006.01.051})

\bibitem{AnHa2021}
Andric, T. and Harms, J.. 2021  {Lightsaber: A Simulator of the Angular Sensing and Control System in LIGO}. {\em Galaxies} \textbf{9}.
(\href{http://dx.doi.org/10.3390/galaxies9030061}{10.3390/galaxies9030061})

\bibitem{MaEA2016}
Martynov, D.V., Hall, E.D. et~al. 2016  {Sensitivity of the Advanced LIGO detectors at the beginning of gravitational wave astronomy}. {\em Phys. Rev. D} \textbf{93}, 112004.
(\href{http://dx.doi.org/10.1103/PhysRevD.93.112004}{10.1103/PhysRevD.93.112004})

\bibitem{IsEA2018}
Isoyama, S., Nakano, H. and Nakamura, T.. 2018  {Multiband gravitational-wave astronomy: Observing binary inspirals with a decihertz detector, B-DECIGO}. {\em Progress of Theoretical and Experimental Physics} \textbf{2018}.
073E01 (\href{http://dx.doi.org/10.1093/ptep/pty078}{10.1093/ptep/pty078})

\bibitem{Harms2021b}
Harms, J., Angelini, L. et~al. 2021  {Lunar Gravitational-wave Antenna}. {\em Decadal Survey on Biological and Physical Sciences Research in Space 2023-2032}.

\bibitem{CoHa2014}
Coughlin, M. and Harms, J.. 2014  {Upper Limit on a Stochastic Background of Gravitational Waves from Seismic Measurements in the Range 0.05 -- 1 Hz}. {\em Phys. Rev. Lett.} \textbf{112}, 101102.
(\href{http://dx.doi.org/10.1103/PhysRevLett.112.101102}{10.1103/PhysRevLett.112.101102})

\bibitem{WaPa1976}
Wagoner, R. and Paik, H.. 1976  Proceedings of International Symposium on Experimental Gravitation, Pavia. Roma Accademia Nazionale dei Lincei, Roma.

\bibitem{BiEA1996}
Bianchi, M., Coccia, E. et~al. 1996  Testing theories of gravity with a spherical gravitational wave detector. {\em Classical and Quantum Gravity} \textbf{13}, 2865--2873.
(\href{http://dx.doi.org/10.1088/0264-9381/13/11/003}{10.1088/0264-9381/13/11/003})

\bibitem{Robson_2019}
Robson, T., Cornish, N.J. and Liu, C.. 2019  The construction and use of LISA sensitivity curves. {\em Classical and Quantum Gravity} \textbf{36}, 105011.
(\href{http://dx.doi.org/10.1088/1361-6382/ab1101}{10.1088/1361-6382/ab1101})

\bibitem{Planck2018}
{Planck Collaboration}, {Aghanim, N.} et~al. 2020  Planck 2018 results - VI. Cosmological parameters. {\em A\&A} \textbf{641}, A6.
(\href{http://dx.doi.org/10.1051/0004-6361/201833910}{10.1051/0004-6361/201833910})

\bibitem{AbEA2017d}
Abbott, B.P., Abbott, R. et~al. 2017  {GW170817: Observation of Gravitational Waves from a Binary Neutron Star Inspiral}. {\em Phys. Rev. Lett.} \textbf{119}, 161101.
(\href{http://dx.doi.org/10.1103/PhysRevLett.119.161101}{10.1103/PhysRevLett.119.161101})

\bibitem{DuEA2022}
Dupletsa, U., Harms, J. et~al. 2022  {GWFish: A simulation software to evaluate parameter-estimation capabilities of gravitational-wave detector networks}. {\em Astronomy and Computing} p. 100671.
(\href{http://dx.doi.org/https://doi.org/10.1016/j.ascom.2022.100671}{https://doi.org/10.1016/j.ascom.2022.100671})

\bibitem{SOFA:2021-01-25}
Board, I.S. IAU SOFA Software Collection. .

\bibitem{astropy:2022}
Collaboration, T.A., Price-Whelan, A.M. et~al. 2022  The Astropy Project: Sustaining and Growing a Community-oriented Open-source Project and the Latest Major Release (v5.0) of the Core Package*. {\em The Astrophysical Journal} \textbf{935}, 167.
(\href{http://dx.doi.org/10.3847/1538-4357/ac7c74}{10.3847/1538-4357/ac7c74})

\bibitem{pyephem2011}
{Rhodes}, B.C.. 2011  {PyEphem: Astronomical Ephemeris for Python}. Astrophysics Source Code Library, record ascl:1112.014.

\end{thebibliography}
\end{document}